\documentclass[letterpaper,twocolumn,10pt]{article}
\usepackage{usenix}

\usepackage{amsfonts}
\usepackage{amsmath}

\usepackage{nicefrac}
\usepackage{siunitx}
\usepackage{array,framed}
\usepackage{booktabs}
\usepackage{
  color,
  float,
  epsfig,
  wrapfig,
  graphics,
  graphicx,
  subcaption
}
\usepackage{textcomp}

\usepackage{setspace}
\usepackage{amsmath}

\usepackage{latexsym,fancyhdr,url}
\usepackage{enumerate}
\usepackage{algorithm2e}
\usepackage{algpseudocode}
\usepackage{graphics}
\usepackage{xparse} %
\usepackage{xspace}
\usepackage{multirow}
\usepackage{csvsimple}
\usepackage{balance}

\DeclareMathOperator*{\argmax}{argmax}

\usepackage{graphicx} 
\usepackage{xspace}
\usepackage{xcolor}
\usepackage{diagbox}
\usepackage{caption}
\usepackage{subcaption}
\usepackage{enumitem}
\usepackage{setspace}
\usepackage{multirow}
\usepackage{booktabs}

\usepackage[leftcaption]{sidecap}
\sidecaptionvpos{figure}{t}

\newtheorem{definition}{Definition}

\newcommand{\figref}[1]{Fig.~\ref{#1}}

\newcommand{\tabref}[1]{Table~\ref{#1}}
\newcommand{\secref}[1]{\S\ref{#1}}

\newcommand{\appref}[1]{App.~\ref{#1}}

\newcommand{\eqnref}[1]{Eq.~\ref{#1}}

\newcommand{\parheading}[1]{\medskip\noindent\textit{\textbf{#1}}\hspace{2pt}}

\newcommand{\ml}[0]{ML\xspace}
\newcommand{\dnn}[0]{DNN\xspace}

\newcommand{\lpnorm}[1]{$\ell_{#1}$\xspace}
\newcommand{\sgd}[0]{SGD\xspace}
\newcommand{\gtsrb}[0]{GTSRB\xspace}
\newcommand{\rgb}[0]{RGB\xspace}
\newcommand{\roa}[0]{$\mathit{ROA}$\xspace}
\newcommand{\doa}[0]{$\mathit{DOA}$\xspace}

\newcommand{\rpattack}[0]{$\mathit{RP}_2$\xspace}
\newcommand{\rpnattack}[1]{$\mathit{RP}_2^{#1}$\xspace}

\newcommand{\abcrown}[0]{$\alpha{}$,$\beta$-crown\xspace}
\newcommand{\tsr}[0]{TSR\xspace}

\newcommand{\instantiation}[0]{\ensuremath{\mathcal{I}}\xspace}
\newcommand{\standard}[0]{\ensuremath{\mathcal{S}}\xspace}
\newcommand{\objstandard}[1]{\ensuremath{s_{#1}}\xspace}
\newcommand{\loss}[0]{\ensuremath{\mathcal{L}}\xspace}
\newcommand{\classifier}[0]{\ensuremath{\mathcal{F}}\xspace}

\newcommand{\advdelta}[0]{\ensuremath{\delta_\mathit{adv}^x}}

\begin{document}

\title{\textsc{Redesigning traffic-signs\\ to Mitigate Machine-Learning Patch Attacks}}

\author{
  {\rm Tsufit Shua, Liron David, Mahmood Sharif}\\
  Tel Aviv University\\
  tsufitronen@mail.tau.ac.il, lirondav@mail.tau.ac.il, mahmoods@tauex.tau.ac.il
}
\maketitle

\begin{abstract}

Traffic-Sign Recognition (\tsr) is a critical safety component for autonomous driving.
Unfortunately, however, past work has highlighted the vulnerability of TSR models to physical-world attacks, through low-cost, easily deployable adversarial patches leading to misclassification. 
To mitigate these threats, most defenses focus on altering the training process or modifying the inference procedure. 
Still, while these approaches improve adversarial robustness, TSR remains susceptible to attacks attaining substantial success rates. 

To further the adversarial robustness of TSR, this work offers a novel approach that redefines traffic-sign designs to create signs that promote robustness while remaining interpretable to humans. 
Our framework takes three inputs: \emph{(1)} A traffic-sign standard along with modifiable features and associated constraints; \emph{(2)} A state-of-the-art adversarial training method; and \emph{(3)} A function for efficiently synthesizing realistic traffic-sign images. 
Using these user-defined inputs, the framework emits an optimized traffic-sign standard such that traffic signs generated per this standard enable training TSR models with increased adversarial robustness.

We evaluate the effectiveness of our framework via a concrete implementation, where we allow modifying the pictograms (i.e., symbols) and colors of traffic signs.
The results show substantial improvements in robustness---with gains of up to 16.33\%--24.58\% in robust accuracy over state-of-the-art methods---while benign accuracy is even improved. 
Importantly, a user study also confirms that the redesigned traffic signs remain easily recognizable and to human observers. 
Overall, the results highlight that carefully redesigning traffic signs can significantly enhance TSR system robustness without compromising human interpretability.

\end{abstract}

\section{Introduction} \label{section:introduction}

Reliable traffic-sign recognition (TSR)---the accurate classification of traffic signs captured by onboard cameras~\cite{STALLKAMP2012323,jia2022fooling,wang2025revisiting}---is vital for the safe operation of autonomous driving (AD) platforms.
As AD technology continues to advance at a rapid
pace, AD vehicles---such as the millions of Tesla vehicles~\cite{Tesla}
currently on public roads---are becoming an increasingly integral part
of everyday life. Ensuring compliance with traffic signs is crucial
for all types of vehicles, whether they are fully autonomous (e.g.,
robo-taxis~\cite{waymo}), semi-autonomous (e.g., Tesla vehicles
equipped with Autopilot~\cite{tesla2}), or traditional human-driven
cars. Failure to obey these rules can result in accidents, posing
significant risks to human safety.

Unlike other perception tasks in AD, such as object and pedestrian
detection~\cite{cao2021invisible, cao2019adversarial}, \tsr{} system  relies mainly on camera
images~\cite{wang2025revisiting}, 
meaning that errors cannot be mitigated by other sensors, such as
LiDAR. As a result, adversarial examples on
image-classification models---i.e., adversarially crafted inputs
aiming to induce misclassification~\cite{athalye2018synthesizing,
  goodfellow2014explaining, kurakin2018adversarial, madry2017towards,
  sharif2016accessorize, szegedy2013intriguing}---pose a distinct 
and serious threat to this task.
Specifically, real-world attacks can be carried out on \tsr by
adding adversarial patches to traffic signs (e.g., in the form of printed
stickers) that are detected by the camera~\cite{eykholt2018robust,
  he2024dorpatch}. 
By leading to the misinterpretation of traffic signs,
these patch-based attacks introduce significant safety risks,
including unexpected emergency braking, speeding violations, and other
hazards.

\begin{figure*}[t!]
  \centering
  \includegraphics[width=0.75\textwidth]{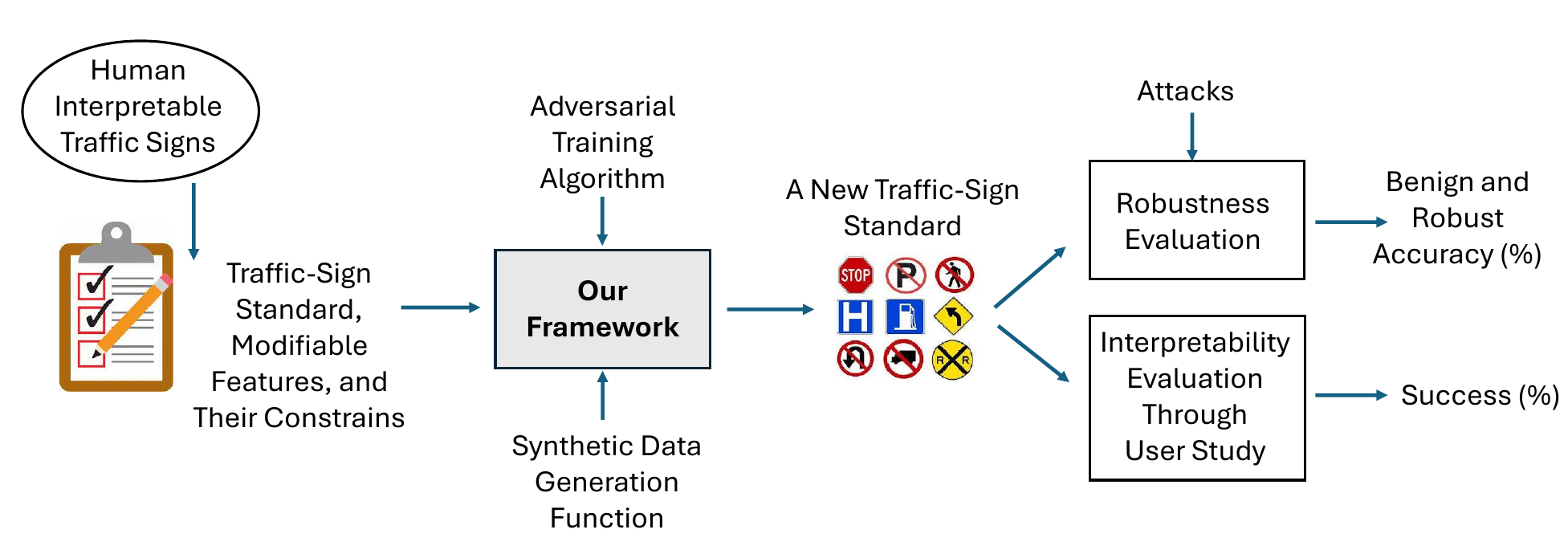}
  \caption{An overview of our suggested framework.
  }
  \label{fig:framework}
\end{figure*}

Several defenses have been proposed against patch attacks. One of the
most effective methods is adversarial
training~\cite{rao2020adversarial, wu2019defending}, which generates
adversarial examples and incorporates them into the training process
to improve model robustness. Still, while it significantly enhances models'
robustness against patch attacks, adversarial training often harms
models' benign accuracy (i.e., accuracy on benign inputs), and models
remain vulnerable to a substantial portion of attack
attempts~\cite{wu2019defending} (e.g., we find that $>$32.47\% of
attack attempts success under various settings in
\secref{sec:eval:results}).
In addition to adversarial training, certifiably robust defenses
(i.e., ones offering provable guarantees) against patch attacks have
been developed~\cite{han2021scalecert, xiang2021patchguard,
  xiang2022patchcleanser, xiang2024patchcure}. However, these
approaches typically assume a restricted threat models---specifically,
that a single, continuous adversarial patch confined to a localized
region is added by the adversary, rendering them vulnerable to
sophisticated attacks~\cite{he2024dorpatch}.

In this paper, we propose a novel alternative approach to enhance the
adversarial robustness of \tsr{} models against a strong adversary capable of distributing adversarial perturbations across multiple patches covering multiple regions, while also improving benign accuracy.
Our approach is based on the insight that the existing human-created
traffic-sign standards can be redefined to improve resilience against
adversarial examples. 
Traffic-sign standards vary across regions, such as the German and
U.S. standards~\cite{serna2018classification}, illustrating that multiple valid
designs already exist. Since these standards are human-defined, we
argue that it is both feasible and beneficial to select or design a
new standard that is not only interpretable by humans but also
optimized to maximize \tsr{} models' reliability.

\subsection{Our Defense Overview} \label{sec:framework}

To find improved traffic-sign standards, we formulate the design process as a robust optimization problem---treating the standard itself as an optimizable variable. 
As depicted in \figref{fig:framework}, our framework takes the following inputs:

\begin{enumerate}

    \item \textbf{A Traffic-Sign Standard, Modifiable Attributes, and Constraints:} A baseline traffic-sign standard, along with a set of attributes that may be slightly modified, and corresponding constraints that ensure these modifications do not alter the sign’s intended meaning or reduce its interpretability for human drivers. These constraints also help maintain compliance with relevant safety standards and regulations, where applicable.
For example, one modifiable attribute could be the color of the pictogram (i.e., the symbol within the sign), with a constraint enforcing that the color remains uniform across the entire pictogram.

    \item \textbf{Adversarial Training Algorithm:} The training process applies adversarial training, incorporating adversarial examples during training, to improve robustness under the evolving traffic-sign design.

As mentioned previously and will be discussed in \secref{sec:defenses}, leading certifiable defenses
against patch attacks, such as PatchCURE~\cite{xiang2024patchcure} and
PatchCleanser~\cite{xiang2022patchcleanser}, are designed to enhance
robustness against a specific class of attacks where the adversarial
patch is continuous and restricted to a single, localized
region. Therefore, these methods are less effective against stronger and
more general adversaries that can distribute perturbations across
multiple, disjoint regions. 

For this broader threat model, adversarial training against patch
attacks~\cite{wu2019defending, rao2020adversarial} remains the
state-of-the-art defense.
Thus, we consider adversarial training on unoptimized standards as
the baseline defense, and incorporate such schemes into the standard
optimization process.

    \item \textbf{Synthesis Data-Generation Function:} Since the modified traffic-signs do not exist in the real world, we use a generator to synthesize images of the redesigned signs for training and evaluation. The dataset generation is repeated numerous times throughout the traffic-sign redesign framework, so it is critical that this process be extremely efficient.
\end{enumerate}

The output of our framework is an optimized traffic-sign standard, where each modifiable attribute is adjusted within its constraints to maximize adversarial robustness. To achieve this, we approach the problem of designing traffic signs that are resilient to patch-based attacks as a robust optimization problem, similar to the adversarial training framework proposed by Madry et al.~\cite{madry2017towards} (\eqnref{eq:PGDRobOpt} in
\secref{sec:background}). While adversarial training focuses on finding model parameters that minimize loss against worst-case perturbations of benign traffic-sign inputs, our goal is to redefine traffic-sign specifications to minimize loss under such perturbations.

Specifically,
we use gradient-based optimization for differentiable attributes, while,
for non-differentiable attributes, we employ a greedy optimization algorithm that selects the best option from a predefined pool of candidates, all of which satisfy the attribute's constraints.
The optimization process iteratively refines the attribute values and evaluates robustness using synthetic data generated from the modified standard.

After obtaining the optimized traffic-sign standard, we evaluate its effectiveness through three key metrics:
\begin{enumerate}
    \item \textbf{Benign  Accuracy:} We assess the model's performance on  benign inputs (without attacks).

    \item \textbf{Robust Accuracy:} We assess the model's performance on adversarial examples by applying the selected state-of-the-art patch-based attacks to traffic signs generated according to the new standard.

    \item \textbf{Human Interpretability:} We conduct a user study to ensure that the redesigned traffic signs remain easily recognizable and interpretable by human observers.
\end{enumerate}

In real-world scenarios, the success of physical adversarial attacks also depends on the resolution and quality of the camera and printer used during fabrication and capture. To ensure that our defense generalizes across a wide range of present and future camera and printer technologies,  we evaluate its robustness under idealized conditions—consistent with the standard practice in most patch defenses~\cite{xiang2022patchcleanser, wu2019defending,xiang2021patchguard,xiang2024patchcure}.

In other words, we consider a \emph{stronger} adversary capable of precisely fabricating adversarial patches without any noise introduced by printing or the camera. While one could evaluate against a particular camera and printer setup, such evaluations are inherently narrow—more advanced equipment could potentially yield higher attack success rates. To account for this, we adopt a worst-case setting where the attacker has access to perfect fabrication and sensing tools. 
This idealized evaluation provides a conservative lower bound on the robustness improvements achieved by our method. In practice, real-world imperfections (e.g., printing artifacts, camera noise) would likely degrade the adversary’s effectiveness, thereby further enhancing the practical robustness of our defense.

\subsection{Our Findings}

To demonstrate the effectiveness of our defense framework, we implement a concrete example in which we chose the modifiable features to be the {pictograms}  (i.e., the symbols within traffic-signs) 
and their {colors}.
The pictogram is constrained to be selected from a predefined set of allowed options, and the color is required to be uniform across all pixels.

For this implementation, we use Defense against Occlusion Attacks
(\doa{})~\cite{wu2019defending} as the adversarial training method (see~\secref{sec:eval:setup:attacks}). The synthetic data generation function we use follows the approach proposed by Maletzky et al.~\cite{Maletzky} (see~\secref{sec:instantiation}), which integrates modified pictograms (including their colors) into realistic traffic-sign images. Specifically, the method replaces the original pictograms in their dataset with the modified versions and renders them in diverse real-world environments.
Since Maletzky et al.~\cite{Maletzky} demonstrated how to generate realistic instances for seven specific traffic signs, we limit our implementation to the same seven signs for compatibility.

Since color is a continuous and differentiable parameter, we apply gradient descent to identify the optimal solid color for the entire pictogram—meaning all pixels share the same color. For pictogram selection, since it is a discrete attribute, we employ a greedy search strategy which chooses the optimal options from a manually predefined pool of pictogram alternatives.

\begin{figure}[t!]
\centering
\includegraphics[width=0.45\textwidth]{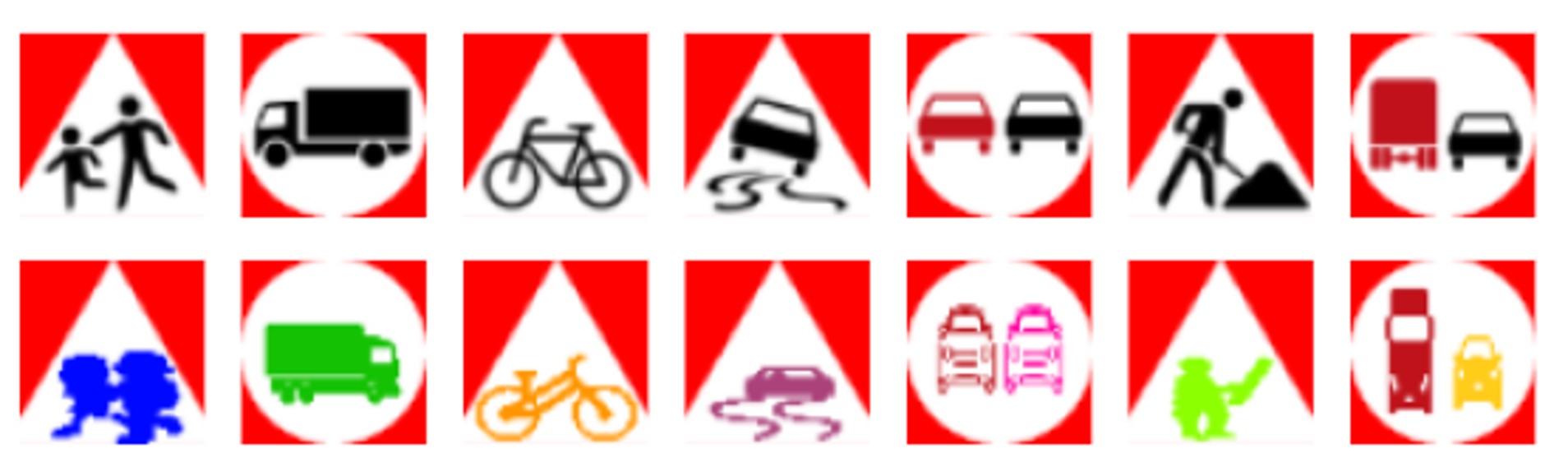}
\caption{
  traffic signs with original pictograms and colors (top)
  vs.\ ones our method creates to defend against
  adversaries in the physical (bottom) domains.}
\label{fig:PictBest}
\end{figure}

We conduct comprehensive experiments using standard image classification architectures, focusing on ResNet-18, which has previously demonstrated strong performance on traffic sign recognition benchmarks~\cite{STALLKAMP2012323, resnet}. We use a potent patch attack ($RP^4_2$, see~\secref{sec:eval:setup:attacks}) out of five we considered, as our main attack, and evaluate models' performance on both benign and adversarially perturbed inputs, reporting both benign accuracy and adversarial robustness. 

Our experimental results show that this concrete implementation improves both
adversarial robustness and benign accuracy.
Specifically, we observe a
significant improvement in adversarial robustness, with up to a
24.58\%
increase in robust accuracy compared to state-of-the-art
techniques. 
Additionally, a user study we conducted demonstrated that human
participants were able to recognize traffic signs generated by our
approach as easily as standard traffic signs. 
\figref{fig:PictBest} shows the redefined traffic signs we obtained.
These results highlight that even minor adjustments to established
standards can substantially boost model reliability. 
We also evaluate our framework under an alternative threat model---specifically, against an adversary introducing imperceptible adversarial perturbations---and observe consistent results (see \appref{sec:addthreatmodel}), further validating the framework's effectiveness.

Next, we provide the relevant background and review related work in~\secref{sec:background}, followed by a description of our threat model in~\secref{sec:threat-model}. We then detail our concrete implementation in~\secref{sec:optimization} and present experimental results in~\secref{sec:eval:results}.
Finally, we conclude the paper in~\secref{sec:conclude}.

\clearpage{}
\newcommand{\eat}[1]{}
\section{Background and Related Work}
\label{sec:background}

\subsection{Traffic-Sign Recognition (\tsr{})  Systems}
To comply with traffic signs encountered on the road, AD vehicles
leverage detection to localize traffic signs and \tsr{} system to classify 
the detected signs by assigning labels to regions in the image containing
traffic signs~\cite{jia2022fooling, wang2025revisiting, STALLKAMP2012323,
  gtsdb_data}.
Detection and classification are typically performed using
deep neural networks (\dnn{}s) (e.g.,~\cite{sermanet2011traffic,
  ren2015faster, chen2019shapeshifter, wu2019defending}) taking
two-dimensional images as inputs~\cite{STALLKAMP2012323, gtsdb_data},
and they may be intertwined~\cite{Redmon2015Yolo} or separate,
occurring one 
after the other~\cite{ren2015faster}.
Similar to \tsr{} (see \secref{sec:back:attacks}), traffic-sign
detection models are also vulnerable against patch-based
attack~\cite{wei2022adversarial,  zhao2019seeing, song2018physical},
calling for means to improve their adversarial robustness.
In line with various past efforts (e.g.,~\cite{cohen2019certified, madry2017towards,
rao2020adversarial, wu2019defending, xiang2022patchcleanser,
  xiang2021patchguard}), however, 
  our work
seeks to improve the adversarial robustness of \tsr{} systems.
Nonetheless, we expect that techniques presented in this work can be
further extended to improve adversarial robustness of traffic-sign
detectors.

\subsection{Patch Attacks}
\label{sec:back:attacks}

\eat{

Various types of attacks against machine learning (\ml) systems exist, each with different goals and capabilities. For example, in training-time attacks, adversaries compromise model accuracy by partially controlling the training data or process~\cite{biggio2018wild, huang2021data}. In privacy attacks, adversaries attempt to extract private information about the training data during or after training~\cite{shokri2017membership}. Availability attacks involve adversaries generating inputs that increase prediction or training time~\cite{shumailov2021sponge}. In contrast, this work focuses on \textit{evasion attacks}, where adversaries can only manipulate inputs during inference to induce misclassifications.

Biggio et al.~\cite{biggio2013evasion} and Szegedy et al.~\cite{Szegedy14AEs} were the first to introduce evasion techniques against machine learning (\ml) models for image classification using \emph{adversarial examples}. To generate these examples, they proposed methods to identify worst-case perturbations, with a small \lpnorm{p}-norm to ensure imperceptibility, that increase the model’s loss and cause misclassification. Since then, numerous studies have proposed more effective, efficient, and imperceptible attacks~\cite{goodfellow2014explaining, carlini2017towards, papernot2017practical, madry2017towards, croce2020reliable}. Typically, adversarial examples are generated by solving an optimization problem of the following form: 
\[
    \argmax_\delta \loss\big(\classifier(x + \delta^x_\mathit{adv}),y\big) \qquad \textit{s.t.}\ \lVert \delta^x_\mathit{adv} \rVert_p \leq \epsilon.
\]
This optimization looks for the adversarial perturbation ($\delta^x_\mathit{adv}$) with bounded \lpnorm{p}-norm (restricted by $\epsilon$), whose addition to the benign sample ($x$) misleads the model (\classifier), by increasing the loss (\loss) where $y$ is the true label associated with the input $x$.

Adversaries differ in their abilities and goals~\cite{Papernot2016Limits}. In \textit{white-box} settings (e.g.,~\cite{madry2017towards}), attackers have full access to the model’s parameters and architecture. In contrast, \textit{black-box} settings (e.g.,~\cite{papernot2017practical}) limit attackers to only querying the model, without access to its internal workings. Attacks can be either targeted, where the goal is to misclassify a specific class, or untargeted, where the objective is any misclassification~\cite{Papernot2016Limits}. Intuitively, untargeted white-box attacks are the most powerful. Therefore, we primarily evaluate the robustness of defenses against such attacks (\secref{sec:threat-model}).

Adversarial perturbations bounded by \lpnorm{p}-norm are well-suited for evaluating robustness in the digital domain, where adversaries have full control over the inputs. However, in real-world scenarios, adversaries often have limited control over inputs (e.g., due to sampling noise). As a result, researchers have also investigated adversarial examples in the physical domain~\cite{sharif2016accessorize, brown2017adversarial, athalye2018synthesizing, eykholt2018robust, he2024dorpatch, zha2020rolma}.
These attacks typically take the form of visible yet inconspicuous perturbations designed to induce misclassification. 

}

Adversarial examples in the physical world have been widely studied~\cite{sharif2016accessorize, brown2017adversarial, athalye2018synthesizing, eykholt2018robust, he2024dorpatch, zha2020rolma}. These attacks typically involve visible yet inconspicuous perturbations designed to induce misclassification by machine learning models. In the context of traffic sign recognition (\tsr), such attacks can be implemented by physically attaching a printed adversarial patch to a traffic sign, which is then captured by the vehicle's camera.

Brown et al.~\cite{brendel2019approximating} introduced one of the earliest physical adversarial patch attacks, demonstrating that a printed patch can reliably cause targeted misclassifications. This result sparked significant interest in the machine learning security community regarding the physical realizability of adversarial attacks. Around the same time, Karmon et al.~\cite{karmon2018lavan} proposed LaVAN (Localized and Visible Adversarial Noise), which focused on patch-like attacks in the digital domain. Since then, numerous patch-based attack methods have been developed to target a range of real-world and simulated scenarios~\cite{eykholt2018robust, he2024dorpatch, liu2018dpatch, song2018physical, wei2022adversarial, wu2020making, zhao2019seeing}.

In this work, we focus on defending against such patch attacks (\secref{sec:threat-model}). Additionally, we show that our approach can be extended to defend against more traditional adversarial perturbations with bounded $\ell_\infty$-norms (\appref{sec:addthreatmodel}).

\eat{
Researchers have investigated adversarial examples in the physical domain~\cite{sharif2016accessorize, brown2017adversarial, athalye2018synthesizing, eykholt2018robust, he2024dorpatch, zha2020rolma}.
These attacks typically take the form of visible yet inconspicuous perturbations designed to induce misclassification. 
Such real-world attacks, in the form of adversarial patches, can be
easily carried out on \tsr by adding a printed patch to a traffic sign
that is detected by the camera. 
Brown et al.~\cite{brendel2019approximating} introduced the first
adversarial patch attack, showing that a physically printed patch can
be used to reliably trigger targeted misclassifications. The physical
realizability of such attacks has since garnered significant attention
within the machine learning security community. In parallel, Karmon et
al.~\cite{karmon2018lavan} proposed a related approach—Localized and
Visible Adversarial Noise (LaVAN) which focuses on digital-domain
patch attacks. Since then, numerous patch-based attack methods have
been developed to target a variety of
scenarios~\cite{eykholt2018robust, he2024dorpatch, liu2018dpatch,
  song2018physical, wei2022adversarial, wu2020making, zhao2019seeing}. 

In this work, we focus on defending against patch attacks
(\secref{sec:threat-model}). We also show that our
approach can be leveraged for defeding against evasive inputs created
through adding adversarial perturbations with bounded
\lpnorm{\infty}-norms to otherwise benign inputs
(\appref{sec:addthreatmodel}).
}

\subsection{Mitigating Patch Attacks}\label{sec:defenses}

Defending \ml{} models against adversarial examples, in general, and
adversarial patches, in particular, is a central challenge in
adversarial \ml{}. 

\subsubsection{Adversarial Training}
One of the most effective techniques for enhancing robustness is
adversarial training, augmenting the training data with correctly
labeled adversarial examples~\cite{goodfellow2014explaining,
  madry2017towards, wu2019defending, pang2022robustness}. Notably,
Madry et al.~\cite{madry2017towards} framed the problem of training
adversarially robust models as a robust optimization problem: 
\begin{equation}
  \min_\classifier \mathop{\mathbb{E}} \Big[ \max_{\delta_\mathit{adv}^x} \mathcal{L}\big( \classifier(x+\delta_\mathit{adv}^x),y \big) \Big].
  \label{eq:PGDRobOpt}
\end{equation}
The objective is to find a classifier (\classifier) that minimizes the
expected maximum loss on adversarial examples. 

Adversarial training has been applied to improve robustness
against adversarial patch attacks~\cite{rao2020adversarial,
  wu2019defending}.   Yet, while improving adversarial robustness,
adversarial training often reduces benign
accuracy~\cite{tsipras2018robustness, stutz2019disentangling,
  wu2019defending}.  
In contrast, we propose a defense that not only achieves
stronger adversarial robustness than standard adversarial training,
but also improves benign accuracy.

\subsubsection{Certifiably Robust Defenses}

Certifiably robust defenses~\cite{chen2022towards,
  chiang2020certified, levine2020randomized, li2022vip,
  mccoyd2020minority, metzen2021efficient} offer provable guarantees
of robustness against adversarial examples. Chiang et
al.~\cite{chiang2020certified} introduced the first certifiably robust
defense for patch attacks using Interval Bound Propagation
(IBP)~\cite{gowal2019scalable, mirman2018differentiable}. Zhang et
al.~\cite{zhang2020clipped} proposed Clipped BagNet (CBN), which
enhances robustness by clipping features from
BagNet~\cite{brendel2019approximating}. 
Levine et al.~\cite{levine2020randomized} developed De-Randomized
Smoothing (DRS), which achieves certified robustness by applying
majority voting over model predictions on multiple small image
crops. DRS has since been extended and improved using Vision
Transformer (ViT)~\cite{dosovitskiy2020image} architectures, as
demonstrated by Smoothed ViT~\cite{salman2022certified},
ECViT~\cite{chen2022towards}, and ViP~\cite{li2022vip}. 

Xiang et al.~\cite{xiang2021patchguard} proposed PatchGuard, a general
defense framework that combines small receptive field (SRF) models for
feature extraction with secure feature aggregation for robust
predictions. Building on this idea,
PatchCURE~\cite{xiang2024patchcure} improved computational efficiency
by further leveraging SRF. 
Xiang et al.~\cite{xiang2022patchcleanser} later introduced
PatchCleanser, which employs a double-masking algorithm to effectively
remove adversarial patches from the input image. Unlike prior
approaches, PatchCleanser does not rely on small receptive field (SRF)
models, yet it achieves state-of-the-art certifiable robustness while
preserving high model utility.

Certifiably robust defenses are tailored to a specific class of
attacks, where the adversarial patch is continuous and confined to a
single, localized region. In contrast, our defense targets a stronger
and more general threat model, capable of spreading perturbations
across multiple, disjoint regions.
As shown in recent work, leading certifiable defenses do not withstand
state-of-the-art attacks under this general threat
model~\cite{he2024dorpatch}.

Several certifiably robust methods have been proposed for attack
detection~\cite{han2021scalecert, mccoyd2020minority,
 xiang2021patchguard}. These approaches seek to detect the presence
of an attack and abstain from making a prediction when one is
identified. While effective in signaling adversarial activity, they
offer a weaker form of robustness compared to defenses that can
recover correct predictions without the need for abstention. 
Our method targets a different threat model and is therefore not
directly comparable to these defenses.

\subsection{Data's Role in Robustness}

Several studies have shown that data distribution plays a critical
role in shaping adversarial robustness~\cite{ilyas2019adversarial,
  Richardson2021Bayes, shafahi2018adversarial,
  shamir2021dimpled}. Building on this insight, we propose a novel
approach to improve robustness against a broader threat model where
adversarial patches may span multiple, disjoint regions, without
compromising benign accuracy. 
Our key idea is to modify the data distribution in \tsr{} to enhance
its robustness to adversarial inputs.

A somewhat related approach by Salman et
al.~\cite{salman2021unadversarial} involved creating a single artifact
that is classified correctly with high confidence across various
imaging conditions, which may not necessarily be adversarial (e.g.,
rain or fog). Similarly, other researchers have explored adding
sample-specific perturbations to enable robust
classification~\cite{Frosio23UnAdv, si23angelic,
  Wang22DefPath}. However, unlike our work, these approaches did not
focus on defending against adversarial
examples~\cite{salman2021unadversarial}, or alter individual objects
(e.g., airplanes) to achieve robustness~\cite{Frosio23UnAdv,
  salman2021unadversarial, si23angelic, Wang22DefPath}. Instead, our
approach aims to define how a traffic-sign standard should be designed
to achieve improved adversarial robustness.

\section{Threat Model}
\label{sec:threat-model}
Our work aims to develop a defense against adversaries who perturb
inputs to induce misclassifications in \dnn{}s used for
\tsr{} systems. Unlike training-time attacks, where adversaries can manipulate
the model's weights or the training process, the adversaries in our
study can \emph{only control the inputs} fed into the deployed
models. 
Specifically, our defense operates by redefining traffic-sign
standards to make \tsr{} less susceptible to such evasion attacks.

We test our defense against \emph{worst-case} adversaries,
following the standard approach in the
literature~\cite{croce2020robustbench, wu2019defending}. Specifically,
we consider adversaries with \emph{white-box} access to \dnn{}s,
aiming to mislead them in an \emph{untargeted} manner. Intuitively,
adversaries with more limited access or those targeting a specific
class would be less effective at circumventing the defense. 

We seek to defend against one of the most well-studied and widely
exploited attack family: adversarial patch aiming to induce
misclassification~\cite{eykholt2018robust, sharif2016accessorize,
  brown2017adversarial,  doan2022tnt, liu2019perceptual,
  bai2021inconspicuous}. 
These attacks can physically interfere with \tsr{} systems by printing
adversarial patterns onto sticker patches and attaching them to
legitimate traffic signs~\cite{eykholt2018robust, he2024dorpatch,
  wei2022adversarial, zhao2019seeing, song2018physical}. In
particular, we consider patches that cover a well-defined
portion of the image (e.g., 5\% of the pixels~\cite{wu2019defending})
to remain relatively inconspicuous.
Yet, in line with the most advanced patch
attacks~\cite{he2024dorpatch}, the adversary may 
introduce multiple patches covering separate regions in the image.
While our defense is designed with patch attacks in mind, we also
demonstrate that it generalizes well to other threat models,
specifically, against adversarial \lpnorm{\infty}-norm-bounded
perturbations (see \appref{sec:addthreatmodel} for more details). 

As mentioned above, and consistent with the standard practice in most patch defenses~\cite{xiang2022patchcleanser, wu2019defending,xiang2021patchguard,xiang2024patchcure}, we assume a stronger adversary that operates under idealized conditions, using perfect cameras and printers, without any fabrication or sensing noise. This setup removes dependence on specific hardware and represents a worst-case scenario. As a result, our measured robustness improvements represent a conservative lower bound—in real-world conditions, where such perfection is unattainable, the actual robustness is expected to be even higher.

\section{Concrete Implementation of Our Defense  Framework}
\label{sec:optimization}

We consider a concrete implementation to demonstrate the effectiveness of our framework as described in \secref{sec:framework}.
We note that while this implementation serves as a concrete example, the framework is flexible, and alternative design choices may yield different standards that also enhance robustness.%

Next, we present our implementation details. Specifically, we first formalize our framework (\secref{sec:formalizing}), then describe the input attributes selected for optimization (\secref{sec:inputs}), and finally outline the efficient and effective optimization processes tailored to these attributes (\secref{sec:opt}).

\subsection{Formalizing Our Framework}
  \label{sec:formalizing}

We formulate the traffic-sign standard design task as an \emph{optimization problem}. To do so, we must first formally define what constitutes a traffic-sign standard, as well as how data can be generated from a given standard during optimization. %

We start with defining the traffic-sign standard:
\medskip
\begin{definition}
 A traffic-sign standard $\standard{}$ defines $n$ classes as a set $\standard{} = \{\objstandard{1}, \objstandard{2}, \ldots, \objstandard{n}\}$, where each $\objstandard{i}$ $(i\in[1,n])$ is a tuple of attributes.
  \label{def:standard}
\end{definition}

Once defined, standards are usually fabricated (many
times) into real-world artifacts, and these artifacts are
placed within different environment. The artifacts are then
sensed (e.g., photographed) under different environmental
conditions (e.g., varied lighting and angles), resulting
in data samples or instances. 
We capture this complex
instantiation process of a standard by the instantiation
operator:
\begin{definition} The instantiation operator
  $\instantiation(\standard)=\{x_1, x_2, \dots\}$ takes as input a
  traffic-sign standard \standard{} and outputs  traffic-sign images
  simulating their appearance in various real-world environments and
  conditions.  
  \label{def:instantiation}
\end{definition}
Said differently, $\instantiation{(\standard{})}$ is the set reflecting the data distribution from which training and test samples are drawn.

The instantiation process is slow, costly, and impractical to perform repeatedly. To address this, we approximate it by synthesizing the dataset using a data generation function. Since this function is used extensively throughout our framework, it is essential that it be highly efficient.

Building on these definitions, we now formulate our optimization objective—to identify a standard $\standard{}$ conducive to adversarial robustness, as well as an adversarially robust model $\classifier$, by:
\begin{equation}
    \min_{\standard, \classifier} \mathop{\mathbb{E}}_{\footnotesize \begin{array}{@{}c@{}} x \sim \instantiation(\objstandard{})\end{array}} \Big[\max_{\delta_{adv}^x} \loss\big(\classifier(x + \advdelta{}), y\big)\Big].
    \label{eq:AllClassOptObj}
\end{equation}
Simply put, this optimization aims to find the traffic-sign standard
(\standard) and model (\classifier) such that, even in the presence of
the worst-case adversarial perturbation (\advdelta), instances
generated according to the traffic-sign standard $\instantiation(\standard)$ would still yield a
low loss (\loss), on average.

Unlike Madry et al.'s adversarial training objective
(Equation~\ref{eq:PGDRobOpt}), which focuses solely on optimizing the
model parameters, our standard optimization
(\eqnref{eq:AllClassOptObj}) jointly optimizes both the
traffic-sign standard and the model. As shown in 
\secref{sec:eval:res:reproduce}, however, once an optimized standard
is fixed, training a model on this standard yields roughly equivalent
robustness.

\begin{figure*}[t!]
  \centering
\includegraphics[width=0.6\textwidth]{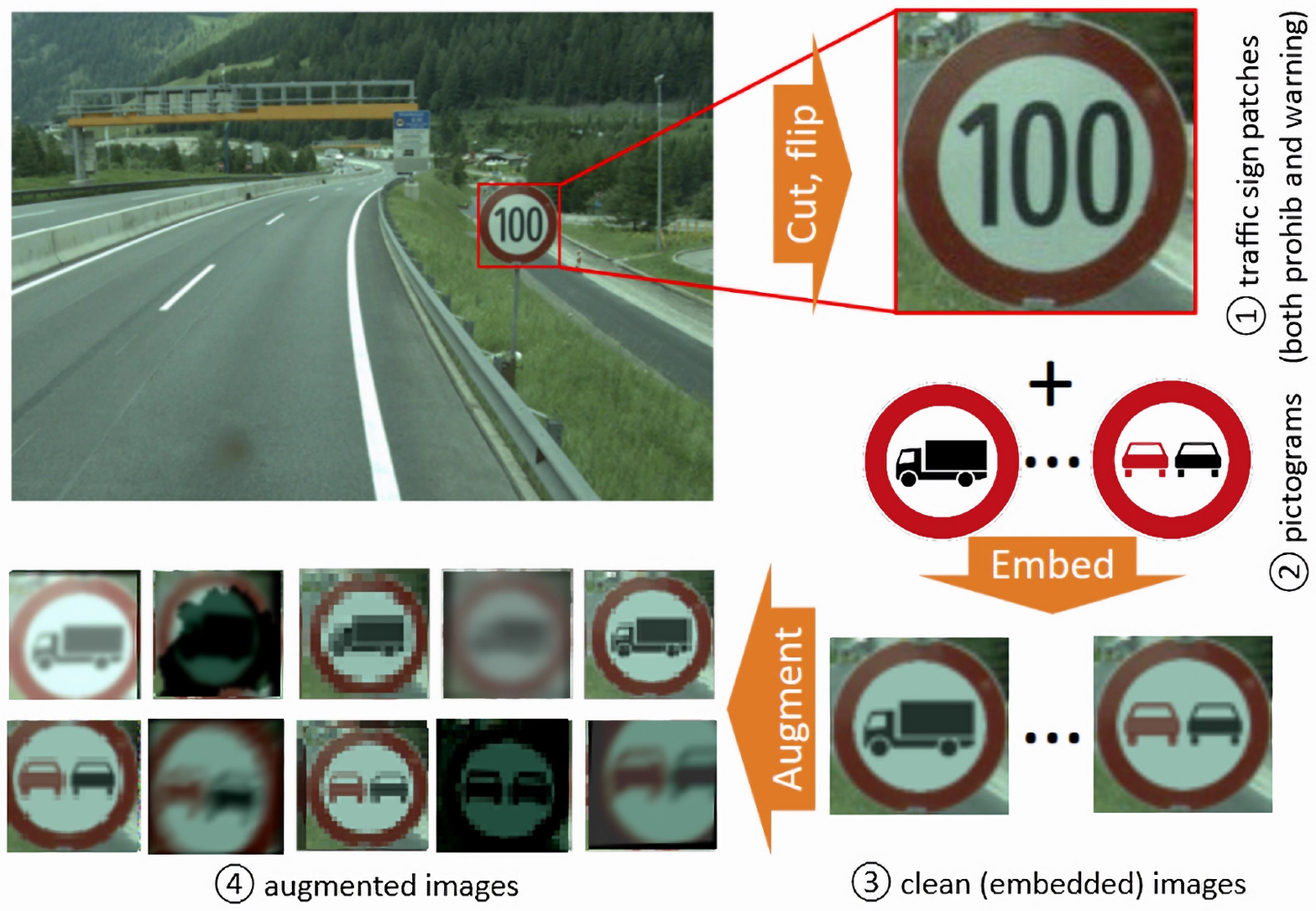}
  \caption{ 
    The dataset synthesis process used for standard instantiation, as presented by Maletzeky et al.\cite{Maletzky}.} 
  \label{fig:SynDataProcess}
\end{figure*}

\subsection{Choosing Concrete Inputs}
\label{sec:inputs}

To solve Equation~\eqref{eq:AllClassOptObj}, we must first model the instantiation process represented by $\instantiation{}$, select the relevant modifiable features (for which we later develop efficient search techniques to jointly optimize $\standard{}$ and $\classifier$), and develop an appropriate optimization strategy. 
We begin by selecting the modifiable features.

\subsubsection{Traffic-Sign Standard and Modifiable Attributes} 

In this work, we adopt the German traffic-sign standard $\standard{}$ as our base and focus on modifying the following key attributes of each class $s_i$:
  \begin{itemize}
      \item $\objstandard{i}[\text{pictogram}]$ representing the
        shape of the pictogram of the traffic sign in class $i$, typically a
        symbol, picture, or number within the sign; 
      \item $\objstandard{i}[\text{color}]$ representing the color of
        the pictogram in class $i$, typically black. 
  \end{itemize}
These attributes are selected due to their direct impact on visual recognition and their feasibility for modification without violating real-world sign conventions.%

We impose the following constraints to ensure both practicality and interpretability:

\begin{itemize}[leftmargin=10pt]
\item Pictogram Shape: Since traffic signs must remain easily interpretable by humans, any modifications to the shape of a pictogram must preserve its original meaning. However, because the shape is inherently non-differentiable, we employ a non-gradient-based optimization approach to explore modifications. To maintain human interpretability during this process, we constrain the optimization to select pictograms from a predefined set of candidate shapes, ensuring that all resulting signs remain recognizable and semantically correct.

\item Pictogram Color: To maintain simplicity and interpretability, we require each pictogram to use a uniform color—that is, all pixels within a pictogram must share the same color. Our goal is to identify, for each selected pictogram, a color that enhances adversarial robustness (measured as increased robust accuracy after shape optimization is complete). Unlike shape, color is a differentiable attribute, allowing us to apply gradient-based optimization techniques to efficiently search for robustness-enhancing color choices.
\end{itemize}

\subsubsection{Synthetic Data Generation Function} \label{sec:instantiation}

To practically insatiate a resulting standard during the optimization process, we synthesize datasets using the method introduced by Maletzeky et al.~\cite{Maletzky}. Their method effectively integrates a modified pictogram (including its color) into real traffic-signs placed within various environments, by replacing the
original pictogram in Maletzeky et al.'s dataset with the modified version.
Maletzeky et al.~\cite{Maletzky} showed how to instantiate main seven traffic signs. Following them, we concentrate on these seven traffic signs in our implementation.

\begin{figure*}[t!]
  \centering
  \includegraphics[width=0.65\textwidth]{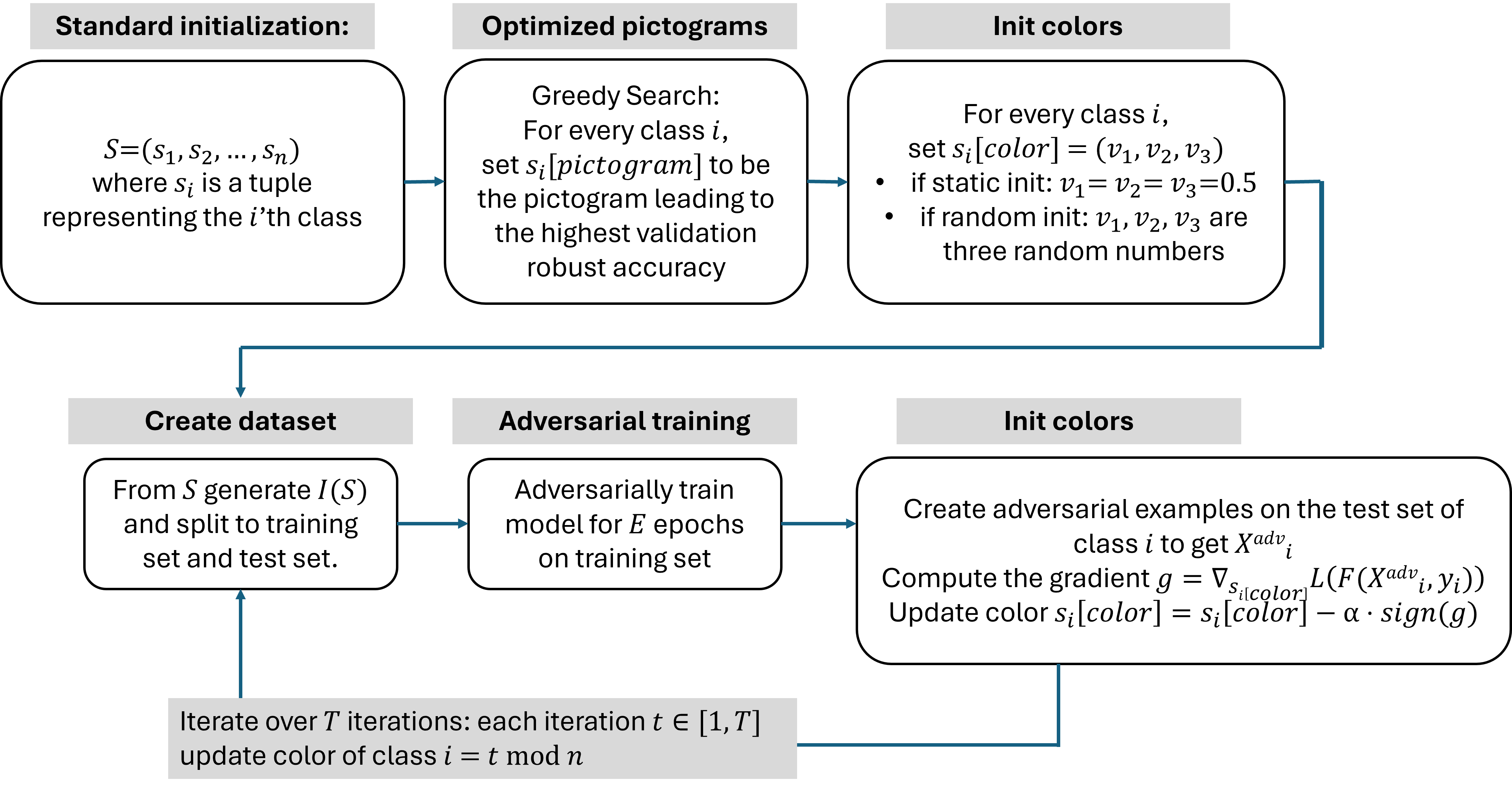}
  \caption{The standard-optimization process. After setting the
    pictograms via greedy optimization and initialing the
    colors according to the selected initialization, we iteratively
    optimize the colors using gradient-based
    optimization.} 
\label{fig:ColorOptProcess}
\end{figure*}

\figref{fig:SynDataProcess} provides an overview
of the entire synthesis process. 
To generate synthetic images in a real-world setting, Maletzky et
al. began by extracting patches containing either a prohibitory sign
(circular with a red border) or a warning sign (triangular with a red
border), referred to as \textit{context scenes}, from high-resolution
photographs of traffic scenes on Austrian highways. 
In total we have 14 context
scenes, seven of these scenes
contain prohibitory signs, while the
other seven contain warning signs. The
number of scenes is doubled through horizontal flipping ({\large
  \textcircled{\small{1}}} in \figref{fig:SynDataProcess}).
Next, each of the seven pictograms is embedded into the appropriate
scenes—both the original and the flipped versions—resulting in an
initial set of 98 images of prohibitory and warning traffic signs
({\large \textcircled{\small{2}}} and {\large \textcircled{\small{3}}} in \figref{fig:SynDataProcess}).

In the next stage, the synthesis method applies image augmentations to
generate realistic traffic-sign images. While Maletzky et
al.~\cite{Maletzky} used the \texttt{imgaug} Python package to
implement augmentations, the transformations in this package are not
differentiable, making them unsuitable for the color gradient-based optimization
process. To address this, we ensure that all augmentation
operations used to generate instantiated images are
differentiable. Specifically, 
we implemented 15 differentiable augmentation methods (such as
Gaussian noise, motion blurring, darkening, etc.) similar to those
used by Maletzky et al., leveraging \texttt{kornia}'s differentiable
operations~\cite{eriba2019kornia}. 
For each of the 98 initial images, we generated 225 augmented images
by applying a set of up to eight randomly selected augmentations with
varying intensities ({\large \textcircled{\small{4}}} in
\figref{fig:SynDataProcess}). 

As a result, the dataset comprised
22,050 images, equally distributed across the source context scenes,
classes, and augmentation methods. 
To assess the generalization ability of the models, we reserved the
images corresponding to two prohibitory-sign and two warning-sign
context scenes for testing, while the remaining images were split into
training (96.7\%) and validation (3.3\%) sets.
Thus, in total, per standard type we instantiated (i.e., with original,
unoptimized signs or optimized ones), we had 15,230 images for training,
520 for validation, and 6,300 for testing.

In \appref{sec:eval:results:instant} we compare the synthesized
dataset with a real-world dataset to qualitatively and quantitatively
evaluate the realism of our traffic-sign synthesis method, showing
that this technique yields realistic images.

\subsubsection{Adversarial Training Algorithm}
\label{sec:eval:setup:training}

We adopted Defense against Occlusion Attacks
(\doa{})~\cite{wu2019defending}, which performs adversarial training
using examples generated by Rectangular
      Occlusion Attack (\roa{} see~\secref{sec:eval:setup}) %
      as it offers open-source
implementation and achieves state-of-the-art defense performance.

We optimized the parameters of the adversarial training method \doa{}
for the best performance. Following the original approach, we first
conducted 30 epochs of standard training using the \sgd optimizer with 
an initial learning rate of 0.03. The learning rate was adjusted using
a step scheduler that decayed it by a factor of 0.1 every 15 
epochs. After this, we performed an additional 100 epochs of
adversarial training, utilizing the Adam optimizer and the cyclic 
learning-rate scheduler~\cite{smith2019super}. 
Similar to the attack phase (\secref{sec:eval:setup:attacks}), during
training, we also used patches covering approximately 5\% of
images' pixels.

\subsection{Optimization Process Based on the Inputs}
\label{sec:opt}

Next, we present a greedy optimization algorithm for selecting the best pictograms, followed by a gradient-based method for determining the optimal colors of the selected pictograms. \figref{fig:ColorOptProcess} illustrates the complete optimization process.

\subsubsection{Pictogram Optimization}
\label{sec:tech:optimization:pictogram}

\begin{figure}[t!]
  \centering
  \includegraphics[width=0.45\textwidth]{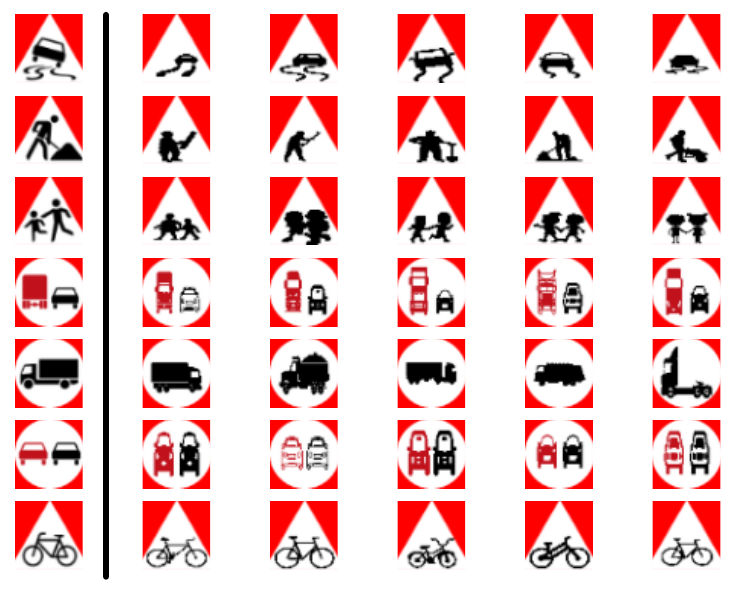}
  \caption{Original pictograms (leftmost column) and the alternatives used in
    pictogram optimization.}
  \label{fig:PictOpt}
\end{figure}

The core idea behind our pictogram shape optimization method is as follows: for each class \( i \in [1, n] \), we construct a candidate pool \( P_i \) consisting of \( m = 5 \) alternative pictograms that are visually similar to the original and maintain human interpretability. To build these pools, we collected five alternatives per traffic-sign class through targeted internet searches (e.g., using keywords like ``bike silhouette''). \figref{fig:PictOpt} shows each original pictogram alongside its corresponding candidate pool. While we used online search for convenience and accessibility, a more curated alternative would be to hire professional designers to create these variants.

Given the pool $P_i$ of each class $i$, ideally we would want to find
the subset
$\{ j_i\}_{i\in[1,n]}$ such that $\{ P_i[j_i] \}_{i\in[1,n]}$
results in the highest robust accuracy. However, finding this optimal
subset is infeasible, so we proceed with a greedy approach.
Specifically, the pictogram optimization process proceeds iteratively,
as outlined below: 
\begin{itemize}
    \item Let $\standard{}$ denote the original traffic-sign standard, where all pictogram colors are black $(0, 0, 0)$. 
    \item Instantiate $\standard$ to obtain $\instantiation{}(\standard{})$. 
    \item For each class $i \in [1, n]$: 
        \begin{itemize} 
            \item For each $j \in [1, m]$: 
                \begin{itemize} 
                    \item Replace the pictogram of class $i$, i.e.,
                      $\objstandard{i}[\text{pictogram}]$, with the
                      $j$-th alternative pictogram $P_i[j]$. 
                    \item Re-instantiate samples to get $\instantiation(\standard)$.
                    \item Perform adversarial training on the model
                      using data from $\instantiation{}(\standard{})$. 
                    \item Evaluate model's robust accuracy to get $a_i$. 
                \end{itemize} 
            \item Set $\objstandard{i}[\text{pictogram}] = P_i[\text{argmax}_{j \in [1, m]} {a_j}]$. 
            \item Re-instantiate samples  to get $\instantiation(\standard)$
              (with the selected pictogram in $\objstandard{i}$).

        \end{itemize} 
\end{itemize}
We considered iterating over the classes in $\standard{}$ in two 
different ways: \emph{(1)} random order, and \emph{(2)} starting with the classes 
with the lowest robust accuracy. We found that the order had little to no 
impact on the results.

\label{sec:eval:setup:optimization:pictogram} 
Due to the stochastic nature of the optimization process, we run each
adversarial training five times, each with a different random
seed. After each run, we compute the corresponding robust accuracy and
calculate the average across the five runs. The pictogram with the
highest average robust accuracy across the five runs for each class
is then selected.

\subsubsection{Color Optimization}
\label{sec:tech:optimization:color}

The inputs to the color-optimization process include:
\emph{(1)} the traffic-sign standard $\standard{}$, where
$\objstandard{i}[\text{pictogram}]$
represents the selected pictogram (as described
in~\secref{sec:tech:optimization:pictogram}) for each $i \in [1,n]$,
and $\objstandard{i}[\text{color}]$ represents the color of each pictogram which is either a
static value (specifically, (0.5, 0.5, 0.5)) or a random value; 
\emph{(2)} the number of iterations $T$; 
\emph{(3)} the number of epochs $E$ for
adversarial training in each iteration; and 
\emph{(4)} a step size $\alpha$ for updating the pictogram colors in 
each iteration. Given these inputs, the optimization process operates
as follows (to ensure the stability of the optimization
process, we update the color of class $i:=t \mod{n}$ 
in each iteration $t\in [1,T]$~\cite{zhu2019deep}): 
\begin{itemize}
\item For each $t\in[1,T]$:
\begin{itemize}
    \item Let $i\gets t\mod{n}$.
    \item Instantiate $\standard{}$ to obtain $\instantiation{}(\standard{})$.
    \item Train the model on $\instantiation{}(\standard{})$ for $E$
      epochs using an adversarial training approach. 
    \item Generate adversarial examples for class $i$ yielding  $X_i^{adv}$.
    \item Compute the (average) gradient of the loss of the adversarial examples $X_i^{adv}$ compared to the their correct label $y_i$, with respect to the $i$\textsuperscript{th} class color $\objstandard{i}[\text{color}]$
    \[ g_i \gets \nabla_{\objstandard{i}[\text{color}]}\loss(\classifier(X_i^{adv},y_i)).\]
    \item Update the color of the $i$\textsuperscript{th} class in the direction opposite to the gradient to minimize the loss
    \[ \objstandard{i}[\text{color}] \gets \objstandard{i}[\text{color}] - \alpha \cdot sign(g_i). \]
    \item Clip the color $\objstandard{i}[\text{color}]$ to ensure it remains within a valid range
    \[ \objstandard{i}[\text{color}] \gets clip(\objstandard{i}[\text{color}],0,1). \]
\end{itemize} 

\end{itemize}

We conducted a line search to determine the optimal parameters (i.e.,
iterations, epochs, and step size) for the color optimization process
that maximized robust accuracy. This resulted in 400 iterations, with
0.25 training epochs per iteration, and a color update step-size of
0.01. 
Given the stochastic nature of the optimization process, we repeat the
color optimization process ten times, each with a different random
seed and a randomly instantiated dataset. For each run, we obtain a
model and a set of optimized $n$ colors. We then select the set of
colors corresponding to the model with the highest robust accuracy.

\section{Evaluation}
\label{sec:eval:results}

We begin by describing the experimental setup used to evaluate both benign and adversarial accuracy (\secref{sec:eval:setup}). Next, we present the results, demonstrating improvements in both benign and adversarial robustness (\secref{sec:ro}). Finally, we assess the human interpretability of new standard through a user study, showing that it preserves interpretability (\secref{sec:eval:results:userstudy}).

\subsection{Experiment Setup}
\label{sec:eval:setup}

\parheading{Model Architectures}
The primary model architecture used in our evaluation is
ResNet-18~\cite{He16ResNet}, which has previously demonstrated high
performance on standard \tsr benchmarks~\cite{STALLKAMP2012323,
  resnet}. To further validate our findings,
we also evaluated models based on other well-known architectures,
such as MobileNet~\cite{sandler2018mobilenetv2},
in \secref{sec:eval:res:reproduce} and \appref{sec:addthreatmodel}.

\begin{figure}[t!]
  \centering
  \includegraphics[width=0.9\columnwidth]{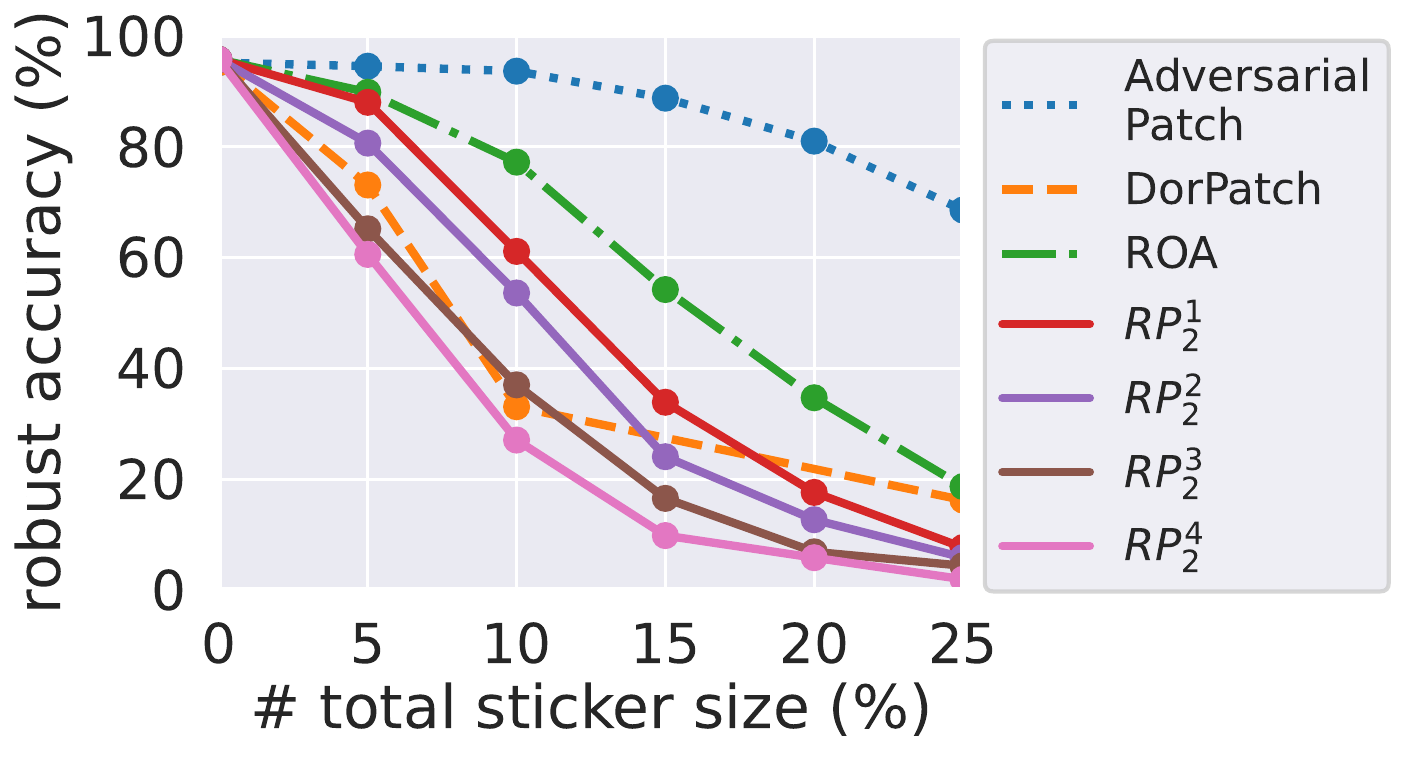}
  \caption{The robust accuracy of a model trained
    via \doa{} against different attacks as a function of the total
    stickers size relative to the input image.} 
  \label{fig:Attacks}
\end{figure}

\begin{figure*}[t!]
  \centering
  \includegraphics[width=0.65\textwidth]{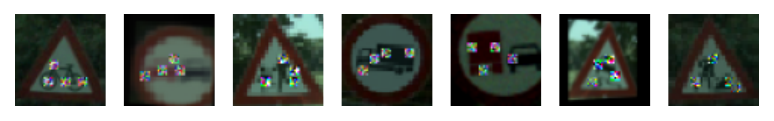}
\caption{Examples of the \rpnattack{4} attack, with patches
    covering $\sim$5\% of the image.} 
  \label{fig:AttackExamples}
\end{figure*}

\parheading{Patch-based Attacks}
\label{sec:eval:setup:attacks}
To choose the best patch attack, we compare the following leading five attacks—four existing methods and one new attack we introduce to strengthen them and choose the strongest among them:
\begin{enumerate} 
    \item Brown et al.'s~\cite{brown2017adversarial}
      \textit{adversarial patch}, which designs a position- and
      input-agnostic (universal) patch that causes misclassification
      into a preselected class.
    \item Wu et al.'s~\cite{wu2019defending} \textit{rectangular
      occlusion attack (ROA)}, which generates image-specific
      adversarial noise in the form of a square, placed in the most
      likely position to induce misclassification.
    \item Eykholt et al.'s~\cite{eykholt2018robust} \textit{Robust
      Physical Perturbations (\rpattack)}, which involves placing
      adversarially crafted black-and-white patcher on signs to
      provoke misclassifications.  
    \item He et al.'s~\cite{he2024dorpatch}
      \textit{DorPatch}, which employs group lasso on patch masks,
      image dropout, density regularization, and structural loss to
      generate multiple adversarial patches       tailored for misleading certified defenses. 
    \item Our \rpnattack{m}, offering an enhancement for \rpattack{}.
      Since the implementation for locating patch positions in Eykholt et
      al.'s attack is not publicly available, we used the fixed locations
      published by the authors for a specific traffic-sign class (namely,
      stop signs). Unfortunately, this led to poor attack-success rates. To
      address this, we combined elements of \rpattack{} and \roa{} to create
      a more effective attack, which we denote as \rpnattack{m}. 
      Similar to \rpattack{}, \rpnattack{m}
      generates multiple patches. Specifically, it identifies $m$
      square-shaped patches, positioning them at locations that
      maximize the loss during initialization (using a fixed color),
      as determined by the exhaustive search approach employed in
      \roa{}~\cite{wu2019defending}. After the locations are set, the
      attack optimizes the patches' colors in the same manner
      as \rpattack{}, but without constraining them to be
      strictly black or white.  
\end{enumerate}
We conducted comprehensive experiments to identify the most effective
attack for evaluating adversarial robustness. For \roa{}, we utilized
the implementation provided by Wu et al.~\cite{wu2019defending}. We
selected the more successful variant, employing an
exhaustive search to position the adversarial patch at the location
that maximizes the loss.
For adversarial patch and \rpattack{}, we also used the implementation
of Wu et al.~\cite{wu2019defending}.
Last, we used DorPatch's official implementation with the default parameters.
However, because its run-time is expensive, requiring more than 12
days to cover the entire test set with its 6,300 samples, we tested
DorPatch on 500 randomly selected samples.
We evaluated the attacks on a model
adversarially trained on the original unoptimized
traffic-sign standard, using
\doa{} (\secref{sec:eval:setup:training}), considering patches of
varying sizes and different values of $m$ in \rpnattack{m}.

\figref{fig:Attacks} presents the results, excluding Eykholt et al.'s
\rpattack{} due to its poor success rates. It is evident that
\rpnattack{4} consistently outperformed all other attacks and
configurations.  
Based on these findings, we selected \rpnattack{4} as the most
effective attack and report the models' robustness accuracy against
it. Specifically we report results for stickers covering approximately
5\% of the image pixels, a scenario commonly studied in the
literature~\cite{wu2019defending}.
Examples of the attacks are shown in \figref{fig:AttackExamples}.

Overall, following our framework, we use the most potent attacks known to date to evaluate the
adversarial robustness of the proposed defense. We emphasize that, as
models' robustness can be enhanced by adversarially training them
directly on optimized traffic-sign standards and the proposed approach
does not introduce randomness or components hindering
differentiability, adaptive evaluation techniques such as backward
pass through differentiable approximation~\cite{athalye2018obfuscated}
are inapplicable.

\subsection{Accuracy and Robustness}
\label{sec:ro}
Our experiments aimed to address several key research questions:

\begin{itemize}

\item \emph{Framework Effectiveness}:
  How does our traffic-sign standard optimization affect both benign
  and robust accuracy? (\secref{sec:eval:res:optim}).

\item \emph{Evaluating Framework Components}:
  To better understand the contribution of each
  component, we evaluated how much the color optimization step
  improves robustness compared to using only pictogram
  optimization. Additionally, we explored the impact of replacing our gradient-based color selection with simpler, more straightforward
  alternatives (\secref{sec:eval:res:baselines}).

\item \emph{Adapting to New Models}:
  In our approach, we apply \eqnref{eq:AllClassOptObj} to
  jointly optimize both the traffic-sign standard and the \tsr{}
  model. Once the traffic-sign standard is determined,
  can we adversarially train robust models
  (e.g., with different or previously unknown architectures)
  \textit{without} reoptimizing the traffic-sign standard? (\secref{sec:eval:res:reproduce}).

\end{itemize}

In this paper, we primarily focus on improving adversarial robustness
of \tsr{} against patch-based attacks. Nonetheless, to assess whether
the standard-optimization approach we propose also applies to other
threat models, we also experiment with defending against adversarial
perturbations bounded in \lpnorm{p}-norm.
Specifically, we repeated all of our experiments to validate our
findings also in the context of widely studied adversarial
perturbations bounded in \lpnorm{\infty}-norm~\cite{croce2020reliable, goodfellow2014explaining, madry2017towards}, where
perturbations per pixel are bounded but can be applied to the entire
image. The results, presented in \appref{sec:addthreatmodel}, closely
mirrored those observed under patch attacks, indicating that our
method generalizes well across different threat models.

\subsubsection{Effectiveness of Standard Optimization}
\label{sec:eval:res:optim}

\begin{figure}[t!]
    \centering
    \includegraphics[width=0.48\textwidth]{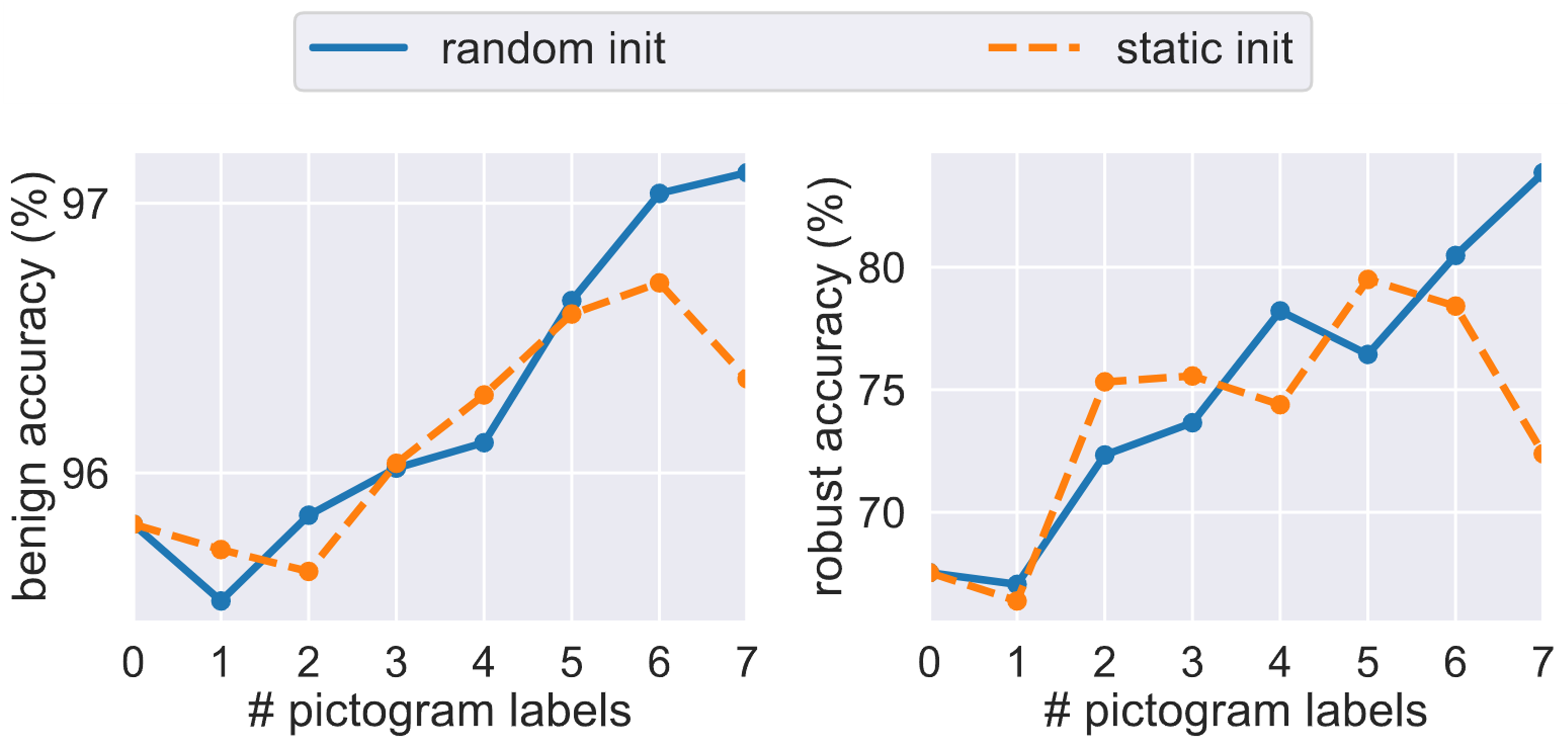}
    \caption{Benign and robust accuracy attained by
      standard optimization, with random and static initialization,
      while increasing the number of optimized pictogram labels. 
      The benign and robust accuracy where the number of optimized
      pictograms labels is 0, corresponds to adversarial training with
      the original (unoptimized) standard.
        } 
\label{fig:StaticVsRandInit}
\end{figure}

To evaluate traffic-sign standard optimization, we began by measuring the benign and robust accuracy of models adversarially trained on the original pictograms and colors. We then progressively increased the number of traffic signs with optimized pictograms and colors using our approach (The optimized standards are shown in \figref{fig:PictBest}.). 
\figref{fig:StaticVsRandInit} presents the results,
comparing two initialization options: one with random color
initialization and one with fixed color values. 
The $x$-axis values represent the following:
0 indicates no optimized traffic signs (i.e., using the original
traffic-sign standard), while 
a value $i \in [1,7]$ indicates that the first $i$ signs are
optimized, with the remaining $7-i$ signs using the original standard.

Compared to adversarial training alone (i.e., 0 optimized pictograms),
optimizing the traffic-sign standard—starting from a random
initialization, led to a substantial improvement in robust accuracy on
the ResNet architecture, with a gain of 16.33\%. This 16.33\% boost in
robustness represents a significant leap, as it far exceeds the
typical gains attained by defenses~\cite{KolterKeynote}.
Additionally, while typical defenses often reduce benign
accuracy~\cite{tsipras2018robustness,stutz2019disentangling}, our
method actually improves it, achieving a 1.30\% gain.  
As can be seen, random initialization generally yielded comparable or
better performance than static initialization, therefore we report the
results using random initialization in subsequent experiments.

\subsubsection{Evaluating Optimization Components}
\label{sec:eval:res:baselines}

Recall that the standard traffic-sign optimization process modifies
both the pictograms and their associated colors. In this section, we
focus specifically on the contribution of color optimization to
robustness, and compare its impact to that of pictogram
optimization. To do so, we compare the performance of pictogram-only
optimization with the additional gains achieved by incorporating color
optimization. 
Moreover, to help further contextualize the results, we investigate
whether simple color-assignment strategies can yield similar
improvements in robustness.

To this end, we evaluate four approaches for selecting traffic-sign
standards. In all cases, pictogram selection follows the approach
described in \secref{sec:tech:optimization:pictogram}, while color
assignment is varied as follows: 

\begin{itemize}[leftmargin=10pt]
\item Default Colors: Each sign retains its original color scheme,
  typically black, or a combination of black and red, without any
  color optimization applied. This condition enables evaluating the
  effect of pictogram optimization alone.

\item Optimized Colors: Our proposed color optimization method, as
  described in \secref{sec:tech:optimization:color}. This condition
  helps in assessing the full standard-optimization process.

\item Random Colors: 
  In this baseline condition, colors are randomly assigned by
  uniformly sampling from the full \rgb{} color space. 

\item Edge (or Extreme) Colors: 
  In yet another baseline condition, we matched colors with signs
  randomly, by permuting the set
  \( C = \{c \in \{0,1\}^3 : c \neq (1,1,1)\} \)
  (consisting of all binary RGB colors excluding white) and assigning
  one color to each traffic-sign class. This strategy reflects the
  tendency of our optimization to favor saturated, high-contrast
  colors near the edges of the \rgb{} space to enhance class
  separability.
\end{itemize}

\begin{figure}[t!]
    \centering
    \includegraphics[width=0.48\textwidth]{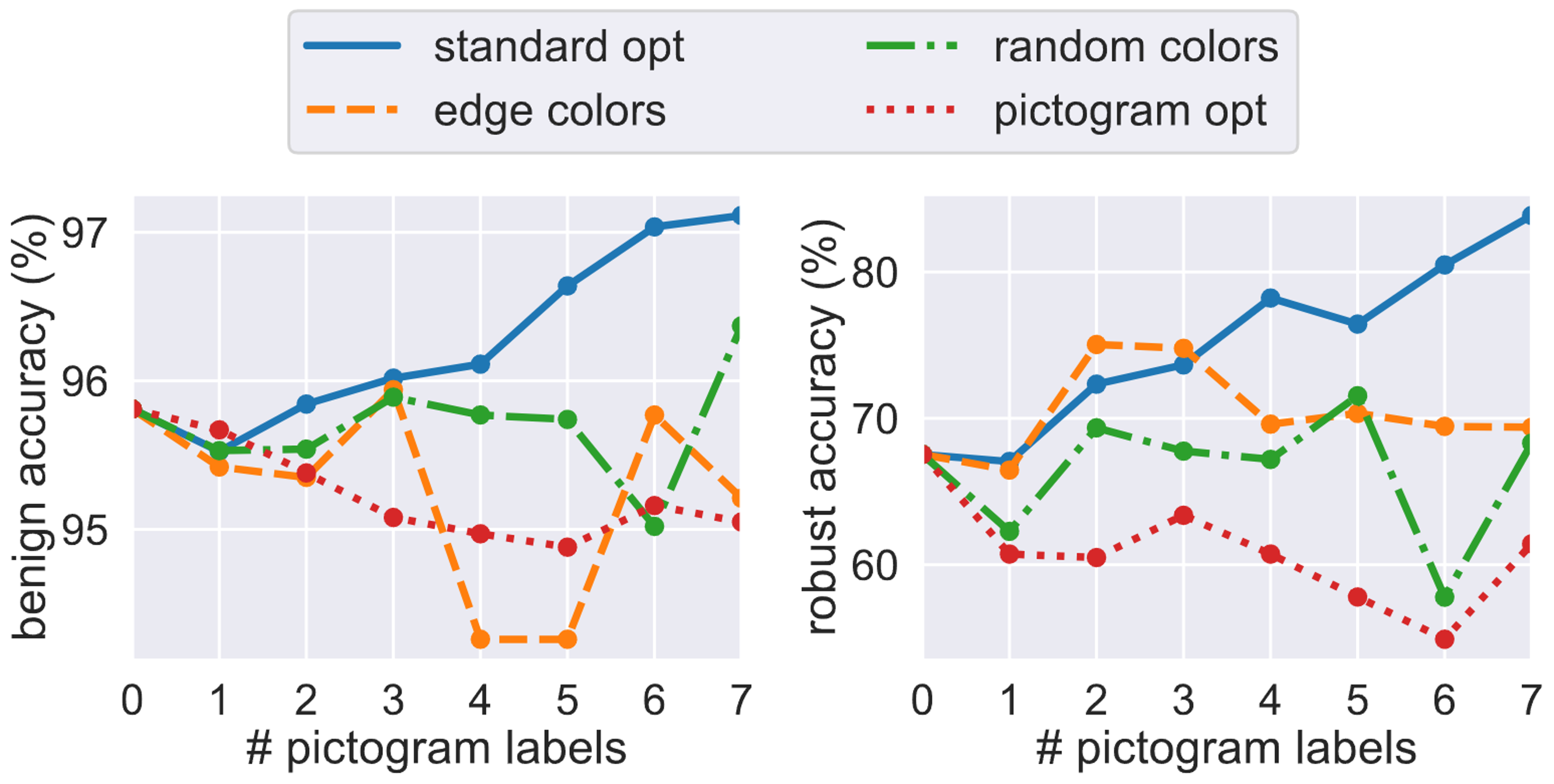}
    \caption{Test benign and robust accuracy of standard optimization
      compared to pictogram optimization alone and setting pictogram
      colors at random or to edge colors.}     \label{fig:OursVsRand}
\end{figure}

The results (\figref{fig:OursVsRand}) show that incorporating our full
standard-optimization process leads to a significant improvements in
both benign and adversarial robustness compared to using
pictogram-optimization alone. While 
pictogram optimization sometimes reduces robustness, the full
traffic-sign standard optimization yields an improvement of $>$15\%.
A similar trend is observed for benign accuracy.

Furthermore, optimizing pictogram colors using our approach resulted
in substantially higher robustness compared to the alternative,
simpler approaches. Both 
random color assignment and selecting distinct edge colors for the
pictograms led to lower robust accuracy than the full traffic-sign
standard optimization when optimizing the colors of all seven
pictograms. These findings highlight the effectiveness of our
traffic-sign standard optimization process, with particular emphasis
on the importance of color optimization.

\subsubsection{Adapting to New Models}
\label{sec:eval:res:reproduce}

Alongside the optimized traffic-sign standard, our optimization
process also produces a model that is jointly trained with the
standard while also optimizing for adversarial robustness.
We now evaluate if it is possible to improve model robustness
by merely adversarially training the model (from scratch) on
traffic-sign standards that have been \emph{a priori} optimized,
without re-running standard optimization.
Moreover, we assess if robustness improves significantly also for
model architectures other than the one used throughout standard
optimization.

\begin{figure}[t!]
    \centering
    \includegraphics[width=0.48\textwidth]{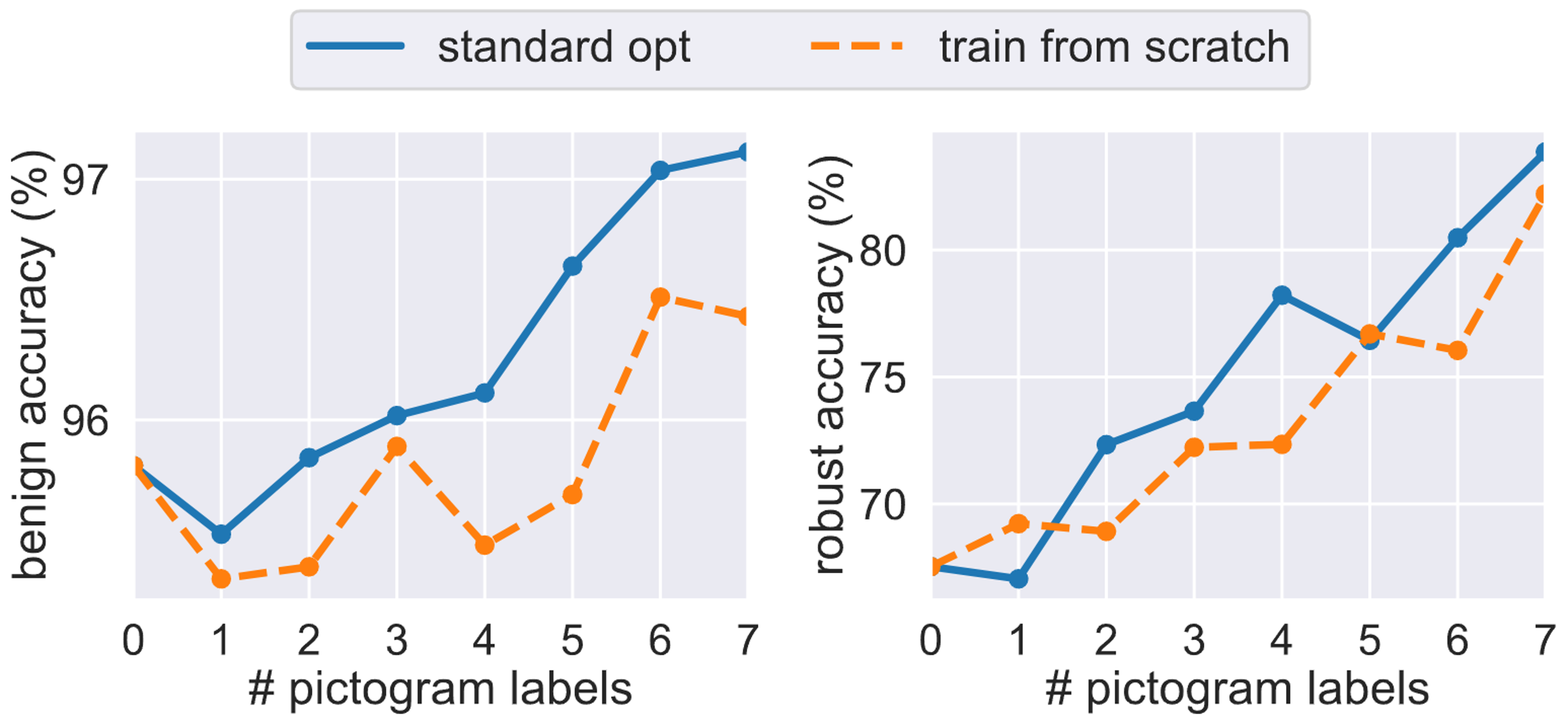}
    \caption{Benign and robust accuracy when training models
      throughout standard optimization vs.\ adversarially training 
      models from scratch using the optimized standards.
    }
    \label{fig:Reprod}
\end{figure}

\begin{figure}[t]
    \centering
\includegraphics[width=0.48\textwidth]{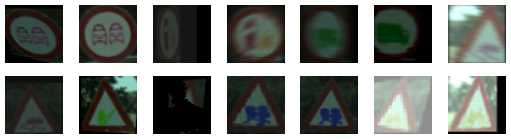}
    \caption{Samples corresponding to the optimized standard that
      participants
      were asked to classify. The samples from the other 
      condition were produced using identical embedding scenes and
      augmentations.}
    \label{fig:study-samples}
\end{figure}
To assess the impact of \emph{a priori} optimized standards on
adversarial robustness, we first adversarially trained ResNet-18
models on datasets instantiated according to traffic-sign standards
derived from running our optimization with varying numbers of pictograms,
without re-running optimization.
\figref{fig:Reprod} presents the results.
It can be seen that the differences in benign and robust accuracy are
negligible between adversarially training the model from scratch on
\emph{a priori} optimized standards vs.\ training the model simultaneously
while optimizing the standard 
(e.g., $<$1.66\% different in robust accuracy when then entire
standard is optimized).

\begin{figure}[t]
    \centering
    \includegraphics[width=0.9\columnwidth]{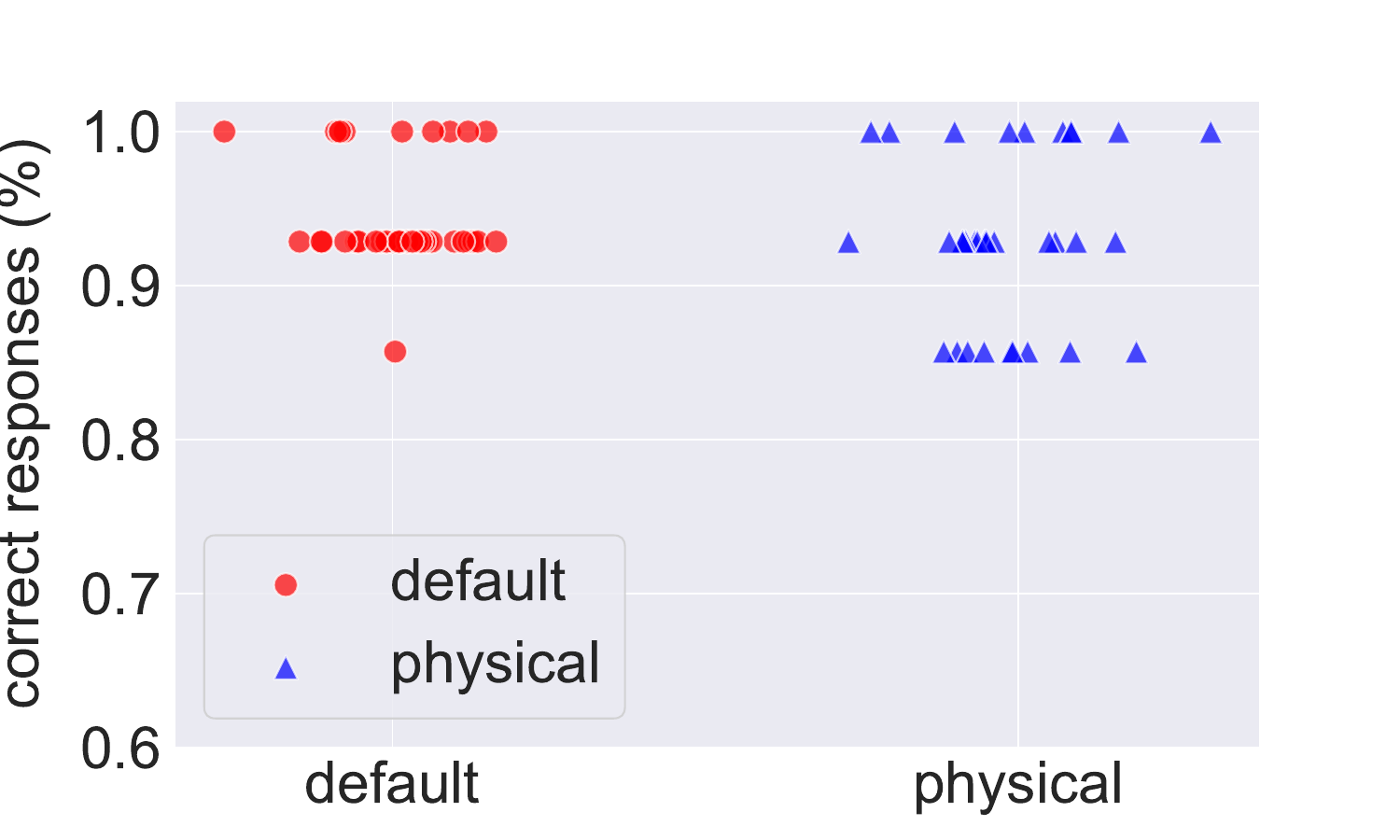}
    \caption{The percentage of correct responses for participants in
      each of the two user-study conditions: default standard (left)
      and optimized standard (right).} 
    \label{fig:scatter}
\end{figure}

Furthermore,
we tested the effect of traffic-sign standard optimization on an
architecture different than the primary ResNet-18 architecture used
throughout our experiments. Specifically, we considered
MobileNet~\cite{sandler2018mobilenetv2}.
Here, we again adversarially trained the model on a standard that has
been optimized \emph{a priori}, for all classes. We then measured the
differences in benign and robust accuracy comapred to adversarial
training on the original, unoptimized standard.
As shown in Table~\ref{tab:Generality}, traffic-sign standard
optimization led to better  performance than adversarial training on
the original standard, yielding higher benign and a significantly higher
robust accuracy, increasing the latter by 24.58\%.

Based on these findings, we conclude that it is possible to
(adversarially) train models on previously optimized standards and
still achieve high levels of adversarial robustness and benign
accuracy. Additionally, standard optimization is also
effective for model architectures other than those considered during
training.

\begin{table}[t!]
    \centering
      \begin{tabular}{l|l|r|r}
        \toprule
        {\bf Architecture} & {\bf Dataset} & {\bf Benign Acc.} & {\bf Robust Acc.} \\
        \midrule
        \multirow{2}{*}{ResNet} &
        Original & 95.81\% & 67.53\% \\
        & Optimized & 97.11\% & 83.86\% \\
        \midrule
        \multirow{2}{*}{MobileNet} &
        Original & 92.50\% & 45.28\% \\
        & Optimized & 95.58\% & 69.86\% \\
        \bottomrule
      \end{tabular}
    \caption{Benign and robust accuracy of models adversarially
      trained and tested using the original traffic-sign standards,
      and on optimized traffic-sign standards, using different
      architectures.}
    \label{tab:Generality}
\end{table}

\begin{figure}[t]
    \centering
    \includegraphics[width=0.65\columnwidth]{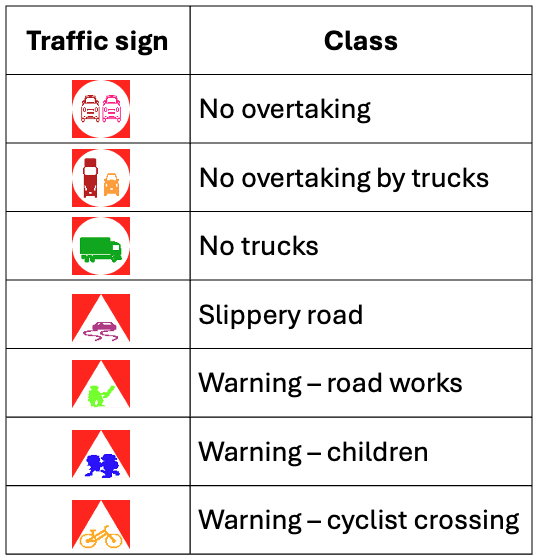}
    \caption{The legend shown at the beginning of the study to
      participants in the optimized condition. An analogous legend 
      was presented to participants in the other condition.} 
    \label{fig:study-legend}
\end{figure}

\subsection{User-Study: Human Interpretability }
\label{sec:eval:results:userstudy}

We designed and ran an online user study to assess whether
human subjects can successfully recognize the artifacts resulting from
the optimized traffic-sign standard.

\parheading{Study Design}
We employed a between-subject design, where each participant was
randomly assigned to one of two conditions and tasked with classifying
traffic-sign images. The two conditions differed based on the
traffic-sign standard used to create the traffic-sign images---one
condition utilized  the default traffic-sign standard, while the other
used the optimized standard. To ensure manageable participation time
and minimize cognitive load, each participant was presented with 14
classification questions, two for each of the seven classes. For
fairness, the 14 images were randomly selected from the respective
synthesized datasets, ensuring that identical embedding scenes and
augmentations were applied across conditions. 

In each of the 14 questions, participants had to choose the correct
class label from seven options in a multiple-choice format. To
mitigate potential ordering effects, we randomized the order of the
questions for each participant. Additionally, we provided a legend at
the beginning of the study to familiarize participants with the
traffic signs. The legend for the optimized condition is shown in
\figref{fig:study-legend}, while \figref{fig:study-samples} presents
the samples participants in the optimized condition had to classify. 

We conducted the study on Qualtrics~\cite{qualtrics} and recruited
participants via Prolific~\cite{prolific}, 
limiting participation to individuals who were at least 18 years old
and fluent in English. Our institution's review board (IRB)
has reviewed and approved our study protocol.

\parheading{Participants} 
A total of 78 participants completed the study. The median completion
time was approximately 3.5 minutes, and participants were compensated
with 1 USD. Four participants were excluded from the analysis due to
anomalous responses—specifically, their classification accuracy was
more than 1.5 times lower than the lowest quartile of the
sample~\cite{seltman12experimental}. As a result, the analysis focused
on the responses of the remaining 74 participants. Participant ages
ranged from 18 to 61 years, with a median age of 29. Of the
participants, 43.5\% self-identified as female, while the remainder
identified as male, which is consistent with a typical Prolific
sample~\cite{peer21MturkVsProlific}. The median self-reported duration
of holding a driving license was 6 years.

\parheading{Study Results}
The scatter plot in \figref{fig:scatter} displays the percentage of
correct responses (i.e., accuracy) for participants in the different
conditions. Participants in the first condition, shown images based on
the default standard, achieved a mean accuracy of 0.94. In comparison,
those presented with traffic-sign images derived from the optimized
standard achieved a mean accuracy of 0.93. A $t$-test revealed that the
difference in mean accuracy between the default and optimized
condition was  \emph{not} statistically significant ($p$-value =
0.11). Therefore, we can conclude that traffic signs optimized using
our approach are roughly as easy to recognize as those based on the
default standards. 

Additionally, we compared the time it took participants to detect the traffic sign under the two conditions to determine if there is a significant difference. To do this, for each condition we first calculated the average detection time per participant. Then, we removed outliers (as in~\cite{seltman2018eda}) %
and computed the overall average detection time using the remaining values.
The results are as follows:
In the default condition (35 participants), the average response time was $t_\text{default}$=15.67 seconds; and,
in the physical condition (36 participants), the average response time was $t_\text{redesign}$=17.57 seconds.
We conducted a one-sided $t$-test with the null-hypothesis $H_0:t_\text{redesign}\ge{}t_\text{default}$. The resulting $p$-value was 0.87, %
indicating that the observed difference in average detection times is not statistically significant, and could easily occur by chance.

\section{Conclusion}
\label{sec:conclude}

Our main goal in this paper is to introduce a novel defense method
for mitigating adversarial examples in \tsr system by redefining a new slightly different traffic-sign standard. 
We first presented our framework then a concrete implementation to demonstrate its effectiveness in practice.

Specifically, we employed gradient-based and greedy optimization techniques to adjust  colors and pictograms attributes, resulting in significant robust accuracy improvements of up to 24.58\%  against adversarial attacks compared to state-of-the-art defenses, while maintaining high accuracy on benign inputs. We then conducted an extensive user study and found that human participants were able to recognize the redesigned traffic signs with high accuracy, indicating that the modified traffic signs remain easily interpretable and suitable for deployment. 
While we present one specific instantiation of the framework, it can be applied in alternative ways to derive different traffic-sign standards that also improve robustness.

\section*{Acknowledgments}

We would like to thank the PLUS research group's members for their
constructive feedback, and Maletzky et al.~\cite{Maletzky} for
sharing their implementation.
This work was supported in part
by a grant from the Blavatnik Interdisciplinary Cyber Research Center
(ICRC);
by Intel\textregistered{} via a Rising Star Faculty Award;
by a gift from KDDI Research;
by Len Blavatnik and the Blavatnik Family foundation;
by a Maof prize for outstanding young scientists;
by the Ministry of Innovation, Science \& Technology, Israel (grant
number 0603870071); 
by a gift from the Neubauer Family foundation;
by NVIDIA via a hardware grant;
by a scholarship from the Shlomo Shmeltzer Institute for Smart
Transportation at Tel Aviv University;
and by a grant from the Tel Aviv University Center for AI and Data
Science (TAD).

\bibliographystyle{abbrv}
\bibliography{cited}

 \appendix
\section{Defending Against Adversarial Perturbations Bounded in \lpnorm{\infty}-norm}
\label{sec:addthreatmodel}

In this section we demonstrate that the proposed traffic-sign standard-optimization method can
apply to other threat models.
To illustrate this, we consider an alternative threat
model where adversaries generate adversarial inputs by adding
perturbations bounded in the \lpnorm{p}-norm to benign
inputs. Specifically, we focus on perturbations bounded in the
\lpnorm{\infty}-norm, which has been extensively studied in the
literature~\cite{croce2020robustbench}.

As for adversarial training, recall that, to select the appropriate
adversarial training, we searched for top-performing adversarial
training techniques that do \emph{not} rely on external unlabeled data
and have open-source implementations. Following this criteria, we used
the \textit{SCORE} adversarial training 
technique~\cite{pang2022robustness}, which ranks amongst the leading
defenses on the RobustBench benchmark~\cite{croce2020robustbench}.
We tuned the adversarial training SCORE parameters for best
performance: per the original work, we used a batch size of 512 and
trained the model for 400 epochs using the SGD optimizer with Nesterov 
momentum~\cite{Nesterov1983AMF} (with a momentum factor of 0.9 and
weight decay 5$\times$10\textsuperscript{-4}). We further used the
cyclic learning rate policy~\cite{smith2019super}, with cosine
annealing of the learning rate from an initial learning rate of
$0.025$ to maximum value of $0.01$ and then to minimal value of
2.5$e$\textsuperscript{-6}.

\begin{figure}[t!]
\centering
\includegraphics[width=0.45\textwidth]{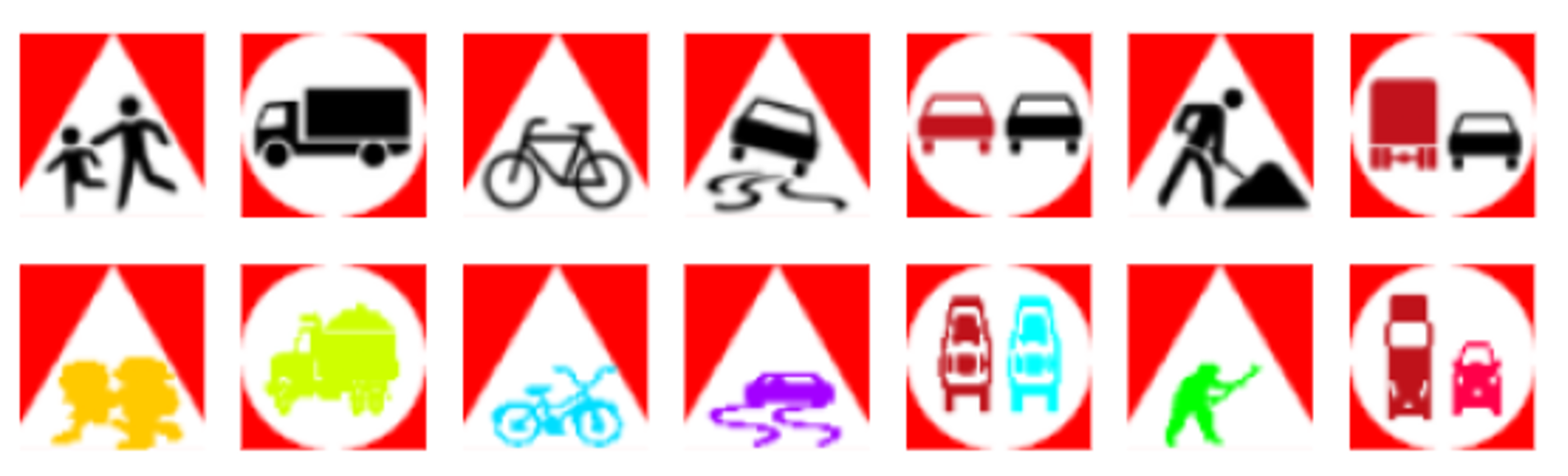}
\caption{
  traffic-signs with original pictograms and colors (top)
  vs.\ ones our method creates to defend against
  adversaries in the alterative \lpnorm{\infty}-norm threat model (bottom).}
\label{fig:PictBest2}
\end{figure}

To evaluate robustness during the optimization process, we used the
AutoAttack~\cite{croce2020reliable}---an ensemble of four, 
advanced attacks, including two white-box attacks, and two black-box
attacks. Notably, for models attaining non-trivial levels of robust
accuracy, the AutoAttack was found to receive empirical robust
accuracy within $<$1\% from more recent, efficient 
attacks~\cite{Liu22Practical, Yao21Automated}. We, however, opted to
use the AutoAttack in our evaluation as it is more commonly used in the
literature~\cite{croce2020robustbench}, enabling a more direct
comparison with previous work. We limited the 
\lpnorm{\infty}-norm of perturbations to $\epsilon = \frac{8}{255}$, as
common~\cite{croce2020robustbench}.
The results are shown in \figref{fig:PictBest2}. 

\subsection{Robust and Accuracy}

Our experiments were designed to address the key research questions
outlined in \secref{sec:eval:results}.
Moreover, we report in~\secref{sec:verification} on our attempts verifying robustness using
state-of-the-art verification tools, providing additional reassurance
on the reliability of our empirical robustness assessment.

\subsubsection*{Effectiveness of Traffic-Sign Standard Optimization}
\begin{figure}[t!]
    \centering
    \includegraphics[width=0.48\textwidth]{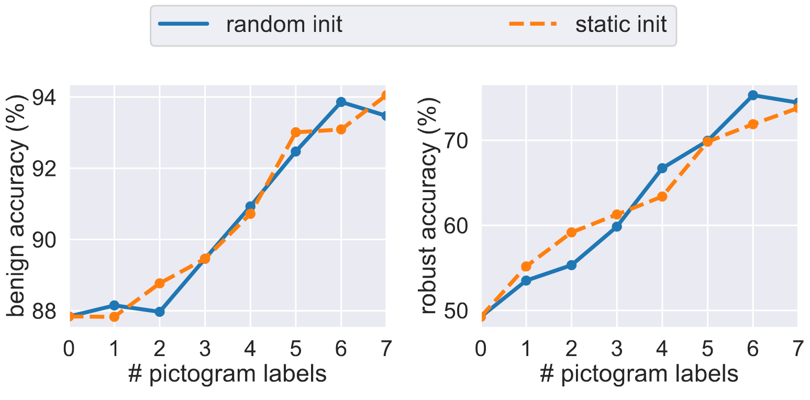}
    \caption{Benign and robust accuracy attained by
      standard optimization, with random and static initialization,
      while increasing the number of pictogram labels optimized,
      under the alternative \lpnorm{\infty}-norm threat model.
      The benign and robust accuracy where the number of pictograms
      labels is 0, corresponds to 
      adversarial training with the original (unoptimized) standard.} 
    \label{fig:StaticVsRandInit2}
\end{figure}

As can be seen in Figure~\ref{fig:StaticVsRandInit2}, compared to the
adversarial training, the trafic-sign standard optimization with
random initialization improved robust accuracy by 25.18\% and an
improvement of 5.63\% was also attained for benign accuracy.  
These results evidence that traffic-sign standard optimization
facilitates adversarial robustness also under this threat model.

We tested whether our results generalize across different
architectures under this threat model too. Specifically, we evaluated
traffic-sign standard optimization on two architectures besides
ResNet: MobileNet and VGG. 
For these architectures, we applied the
optimization across all seven classes and measured both benign and
robust accuracy, comparing the results with the performance achieved
through adversarial training on the original standard.

As shown in \tabref{tab:Generality1}, our results generalize across
architectures. Traffic-sign standard optimization consistently
achieved higher benign and robust accuracy than adversarial training
on all architectures under this threat model as well.

\begin{table}[t!]
    \centering
    \begin{subtable}[t!]{0.5\textwidth}
      \centering
      \resizebox{\textwidth}{!}{
      \begin{tabular}{l|l|r|r}
        \toprule
        {\bf Architecture} & {\bf Dataset} & {\bf Benign Acc.} & {\bf Robust Acc.} \\
        \midrule
        \multirow{2}{*}{ResNet} &
        Original & 87.84\% & 49.25\% \\
        & Optimized & 93.47\% & 74.43\% \\
        \midrule
        \multirow{2}{*}{MobileNet} &
        Original & 77.29\% & 28.09\% \\
        & Optimized & 89.33\% & 60.58\% \\
        \midrule
        \multirow{2}{*}{VGG} &
        Original & 87.98\% & 48.55\% \\
        & Optimized & 93.34\% & 73.88\% \\
        \bottomrule
      \end{tabular}
      }
    \end{subtable}
    \newline
    \caption{Benign and robust accuracy of models adversarially
      trained and tested using the original standards, and on
      optimized standards, using different architectures.}
    \label{tab:Generality1}
\end{table}

\subsubsection*{Evaluating Traffic-Sign Optimization Components}
\begin{figure}[t!]
    \centering
    \includegraphics[width=0.48\textwidth]{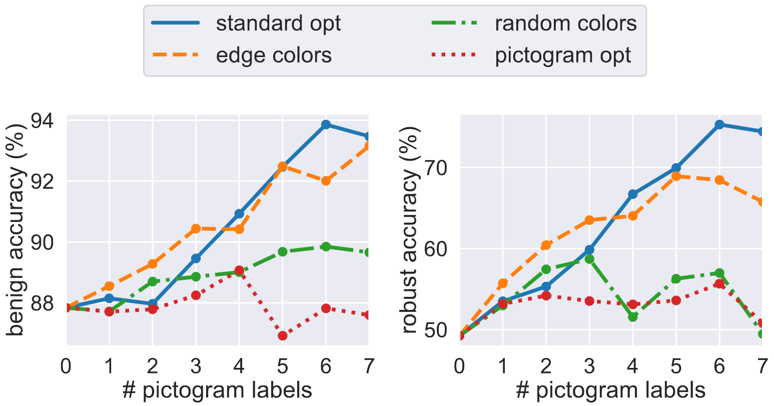}
    \caption{Test benign and robust accuracy of standard optimization
      compared to pictogram optimization alone and setting pictogram
      colors at random or to edge colors, under the alterative \lpnorm{\infty}-norm threat model.} 
    \label{fig:OursVsRand2}
\end{figure}

The results (\figref{fig:OursVsRand2}) show that the full traffic-sign
standard optimization, which includes both pictogram and color
optimization, significantly outperforms pictogram optimization alone, achieving an improvement of over 20\% in robust accuracy and over 5\% in benign accuracy. 

Additionally, optimizing pictogram colors using our approach resulted
in considerably higher robustness compared to the baselines under this
threat model too. Both random color assignment and selecting distinct
edge colors for the pictograms yielded lower robust accuracy than the full traffic-sign standard optimization when optimizing  the colors of all seven pictograms.

\subsubsection*{Adapting to New Models}
\begin{figure}[t!]
    \centering
    \includegraphics[width=0.48\textwidth]{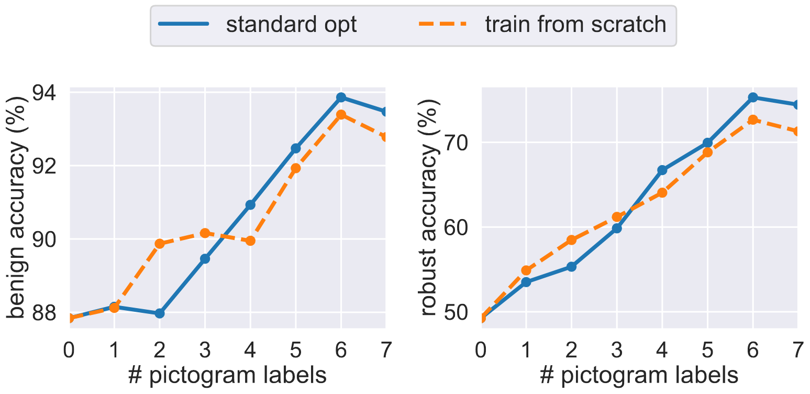}
    \caption{Benign and robust accuracy when training models
      throughout standard optimization vs.\ adversarially training
      models from scratch using the optimized standards, 
      under the alternative \lpnorm{\infty}-norm threat model. 
    }
    \label{fig:Reprod2}
\end{figure}

The difference in robust accuracy between models trained from scratch
and those obtained through the optimization process was small
($\pm$2\%)  as can be seen in \figref{fig:Reprod2}. Therefore, we conclude that adversarially training models from scratch using the optimized traffic-sign standard under this
threat model can effectively replicate the results of the traffic-sign standard optimization process.

We also assess whether traffic-sign standard optimization can enhance robust accuracy when training models with techniques outside those
used in the standard optimization process. To do this,  we conducted the following
experiment: 

We started by applying adversarial training with SCORE to
optimize the traffic-sign standard. Next, we adversarially trained
models from scratch on both the optimized and the original unoptimized
traffic-sign standard, using an alternative adversarial training
method based on perturbations bounded in the \lpnorm{\infty}-norm,
rather than SCORE. Specifically, we used DAJAT, a high-performing
adversarial training method~\cite{croce2020robustbench}, which
leverages advanced data augmentations to enhance adversarial
robustness while preserving benign accuracy. We employed the official
implementation provided by the authors and trained ResNet-18 models
using DAJAT's default parameters. After training, we evaluated benign
accuracy on the respective test sets of the two datasets and ran the
AutoAttack to measure robustness.

The results obtained with
DAJAT were consistent with those from SCORE. The model trained on
samples based on the optimized traffic-sign standard achieved 69.26\%
robust accuracy, compared to 52.67\% robust accuracy for the model
trained on samples from the default traffic-sign standard. This
represents a 16.59\% improvement in robust accuracy due to standard
optimization. Additionally, we observed an increase in benign accuracy
following standard optimization (92.59\% vs. 89.80\%). These findings
demonstrate that traffic-sign standard optimization can enhance robust
accuracy even when training models with techniques other than those
incorporated into the standard optimization process.

\subsection{Human Interpretability}
The study design is similar to that described in
\secref{sec:eval:results:userstudy}, with one key difference: the
optimized traffic-signs shown to the participants were based on the
threat model involving perturbations bounded in the  
\lpnorm{\infty}-norm, rather than 5\% patch cover.

Participants in the first condition, presented with images based on the default standard, achieved a mean accuracy of 0.94. In comparison, participants shown traffic-sign images based on the optimized standard achieved a mean accuracy of 0.90.
A t-test revealed that the difference in mean accuracy between the two
conditions was statistically significant ($p$-value $<$ 0.01), though the effect size was small (0.04).

Additionally, in the default condition (35 participants), the average response time was 15.67 seconds, while in the other condition (38 participants), the average response time was 17.97 seconds.
The resulting p-value is 0.8961, indicating that the observed difference in average detection times is not statistically significant and could easily occur by chance.

\section{Additional Validation} \label{sec:valid}

We now present several additional experiments to further validate our
findings and methodology. Specifically,
in~\secref{sec:verification} we show  robustness verification, 
in \secref{sec:eval:results:instant} we compare the synthesized
dataset with a real-world dataset to evaluate the realism of our
traffic-sign synthesis method, 
and in \secref{sec:eval:results:time} we provide run-time measurements.

\subsection{Robustness Verification}
\label{sec:verification}
To establish a lower bound on robust accuracy, we attempted to use
\dnn{}-verification tools to confirm that no misclassified adversarial
examples could be generated where attacks failed. To this end, we
employed \abcrown{}~\cite{Kotha23ABCrown}, a state-of-the-art
verification well-suited for verifying adversarial robustness. While
\abcrown{} is sound, it is incomplete. In other words, if the
verification results indicate that no adversarial examples exist under
the given constraints, then none can be created. However, since \dnn{}
verification is NP-hard, verification tools may sometimes fail to
confirm robustness within a finite time frame, even if no
misclassified adversarial examples can be generated.

We executed \abcrown{} for samples where AutoAttack
failed to produce adversarial examples against either the ResNet or
VGG models on the optimized standard. %
(MobileNet does
not lend itself for verification with \abcrown{} due to incompatible
layers.) For the VGG model, we found that \abcrown{} ran out of memory
even with our largest GPU (NVIDIA A100 with 80GB or memory). For
ResNet, \abcrown{} did not find any counterexamples (i.e., samples
that could be misclassified), but also failed to verify robustness for
any sample, even when setting the wall-time to a permissive value of
10 minutes per sample. These results held for different
parameterizations of \abcrown{}.

We can conclude that established verification tools may be
insufficiently mature for verifying robustness under the threat
models, \dnn{}s, and datasets we consider. However, \abcrown{}'s
failure to demonstrate misclassified samples after a long run time
provides further validation of our empirical robustness evaluation
with the AutoAttack.

\subsection{Standard Instantiation}
\label{sec:eval:results:instant}
We aimed for our synthesized dataset to closely resemble real-world
traffic-sign datasets. To achieve this, we used the German \tsr
Benchmark (\gtsrb)~\cite{STALLKAMP2012323}, a standard dataset
containing 48$\times$48 images, as a reference. We leveraged it to
both qualitatively and quantitatively assess the similarity between
our synthetically generated data and real data. For this comparison,
we used all \gtsrb{} samples from the same seven classes included in
our synthesized dataset.

\begin{figure}[t!]
  \centering
  \includegraphics[width=\columnwidth]{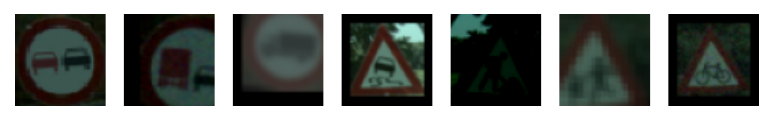}\\
  \includegraphics[width=\columnwidth]{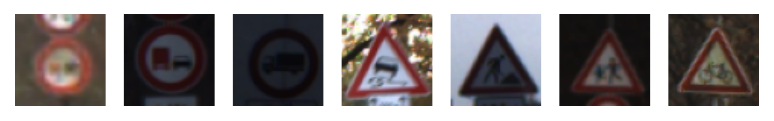}
  \caption{Examples from the synthetic dataset (top) vs.\ examples
    from the corresponding \gtsrb classes (bottom).}
  \label{fig:GTSRBVsSynthetic}
\end{figure}

To qualitatively evaluate the synthesized samples, we visually
compared them with images from \gtsrb{}. \figref{fig:GTSRBVsSynthetic}
shows examples from both the synthetic dataset and \gtsrb{}. As
observed, the images from the \gtsrb{} dataset and the synthesized
samples closely resemble each other, making them nearly
indistinguishable."

In addition to the qualitative assessment, we quantitatively measured
the similarity between our synthesized dataset and \gtsrb{} by
training models on one dataset and evaluating them on the
other. \tabref{tab:GTSRBVsSynthetic} summarizes the results. We
observe that the benign accuracy of the model trained and tested on
the synthetic dataset is only slightly lower (approximately 4\%) than
that of the model trained and tested on samples from the corresponding
seven classes of \gtsrb. 

For the models trained on the synthetic dataset using adversarial
training (with \doa{} as described in
\secref{sec:eval:setup:training}), the benign accuracy when tested on
the synthetic dataset is even higher (approximately 2\%) than that of
the standard training. When tested on \gtsrb, the model trained on the
synthetic dataset experiences a decline of up to 14\% in benign
accuracy, as expected. However, with adversarial training, this drop
is limited to only 2\%. 
Overall, the results suggest that the synthetic dataset exhibits
characteristics similar to those of \gtsrb{}. This indicates that the
findings from our experiments on the synthesized dataset are likely to
generalize to real-world scenarios.

\newcolumntype{P}[1]{>{\centering\arraybackslash}p{#1}}
\begin{table}[t!]
\centering
\begin{tabular}{l|l|r|r} 
  \toprule
\multirow{2}{0.1\textwidth}{\bf \centering Training Dataset} &
\multirow{2}{0.1\textwidth}{\bf \centering Training Method} &
\multicolumn{2}{c}{\bf Test Dataset} \\
& & {\bf Synthetic} & {\bf \gtsrb}\\ \midrule
\gtsrb & Standard & ----- & 97.18\% \\ \midrule
\multirow{3}{*}{Synthetic} & Standard & 93.31\% & 79.49\% \\ \cmidrule{2-4}
& Adv. & 95.81\% & 93.38\%\\ \bottomrule
\end{tabular}
\caption{Comparison between benign accuracy of a model trained and
  tested on \gtsrb and models trained (via standard and adversarial
  training) on the synthetic dataset and tested with the synthetic
  dataset and \gtsrb.  
}
\label{tab:GTSRBVsSynthetic}
\end{table}

\subsection{Run-Time Measurements}
\label{sec:eval:results:time}

To measure run times, we performed adversarial training and standard
optimization on a single GPU, running the process once on an NVIDIA
A5000 and once on an NVIDIA GeForce RTX 3090. The average run time for
adversarial training was 8.17 hours.  

Standard optimization, which includes adversarial model training as
part of the process (see \secref{sec:optimization}), took longer
but still completed in a manageable time, requiring 22.83 hours.  
While standard optimization took longer, we note that we have not
specifically optimized for time efficiency, so significant
improvements to the run time are possible. Additionally, as
highlighted in our evaluation (\secref{sec:eval:res:reproduce}), once
an optimized standard conducive to adversarial robustness is
established, models can be trained more efficiently through
adversarial training alone on the optimized standard
(without re-running standard optimization).

\end{document}